\documentclass[sts,preprint]{imsart}

\RequirePackage[OT1]{fontenc}
\usepackage{amsthm,amsmath,natbib}
\RequirePackage[colorlinks,citecolor=blue,urlcolor=blue]{hyperref}

\usepackage{amsfonts,graphicx,multirow,bm,tabularx,xypic}%,subcaption,caption}
\newcommand{\ERmodel}{Erd\H{o}s-R\'{e}nyi}
\newcommand{\BAmodel}{Barab\'{a}si-Albert}

\startlocaldefs
\numberwithin{equation}{section}
\theoremstyle{plain}

\endlocaldefs

\begin{document}

\begin{frontmatter}
\title{The Geometry of Continuous Latent Space Models for Network Data}
\runtitle{The Geometry of CLS Models for Network Data}

\begin{aug}
\author{\fnms{Anna L.} \snm{Smith}\ead[label=e1]{als2356@columbia.edu}},
\author{\fnms{Dena M.} \snm{Asta}\ead[label=e2]{dasta@stat.osu.edu}}
\and
\author{\fnms{Catherine A.} \snm{Calder}\ead[label=e3]{calder@stat.osu.edu}
\ead[label=u1,url]{www.foo.com}}

%\thankstext{t1}{Columbia University}
%\thankstext{t2}{The Ohio State University}
\runauthor{A. Smith, D. Asta and C. Calder}

%\affiliation{Columbia University and The Ohio State University}

\address{Anna L. Smith is a Postdoctoral Researcher at the Department of Statistics, Columbia University, 1255 Amsterdam Avenue, New York, New York 10027 \printead{e1}.}
\address{Dena M. Asta is an Assistant Professor at the Department of Statistics, The Ohio State University, 1958 Neil Avenue, Columbus, Ohio 43210 \printead{e2}.}
\address{Catherine A. Calder is a Professor at the Department of Statistics, The Ohio State University, 1958 Neil Avenue, Columbus, Ohio 43210 \printead{e3}.}
\end{aug}

\begin{abstract}
% 200 words or less
We review the class of continuous latent space (statistical) models for network data, paying particular attention to the role of the geometry of the latent space. In these models, the presence/absence of network dyadic ties are assumed to be conditionally independent given the dyads' unobserved positions in a latent space. In this way, these models provide a probabilistic framework for embedding network nodes in a continuous space equipped with a geometry that facilitates the description of dependence between random dyadic ties.   Specifically, these models naturally capture homophilous tendencies and triadic clustering, among other common properties of observed networks.  In addition to reviewing the literature on continuous latent space models from a geometric perspective, we highlight the important role the geometry of the latent space plays on properties of networks arising from these models via intuition and simulation.   Finally, we discuss results from spectral graph theory that allow us to explore the role of the geometry of the latent space, independent of network size.  We conclude with conjectures about how these results might be used to infer the appropriate latent space geometry from observed networks.
 \end{abstract}

\begin{keyword}
\kwd{geometric curvature}
\kwd{graph Laplacian}
\kwd{latent variable}
\kwd{network model}
\end{keyword}

\end{frontmatter}

%%%%%%%%%%%%%%%%%%%%%%%%%%%%%%%%%%
%%
%%    (  1  )   INTRODUCTION
%%
%%%%%%%%%%%%%%%%%%%%%%%%%%%%%%%%%%

\section{Understanding network data through geometric embeddings} \label{secIntro}
An important consideration in the development of an analytic strategy for understanding network data is the expected complexity of the types or patterns of connections in the network.  In a statistical framework, we can formalize this notion through a particular dependence structure among the potential ties in the network \citep{kolaczyk_csardi_2014,airoldi_blei_fienberg_etal_2008book}.  For example, a simple statistical model might treat each pair of nodes, or dyad, independently \citep{erdos_renyi_1960,gilbert_1959}.  Of course, this is typically not a reasonable assumption.  For example, in a social network, it as (at least heuristically) reasonable to assume that ``the friend of my friend is also a friend of mine'' \citep{rapoport_1953}.  Indeed, this transitivity effect has strong theoretical support in the social network setting \citep{simmel_1950, holland_leinhardt_1970, coleman_1990} and has been observed in a variety of empirically observed networks as well \citep{davis_1970, goodreau_kitts_morris_2009, watts_1999}.  This expectation for complex non-dyadic dependence among nodes is not unique to social networks or the social sciences, but instead permeates network applications from various disciplines.  However, note that such complex interactions directly contradict an assumption of dyad independence.  If there is a connection between the $i$th and $j$th node and a connection between the $j$th and $k$th node, this tells us something about the probability of a connection between the $i$th and $k$th node.

This of course raises a natural question:  What type of dependence is appropriate and how might this dependence affect an analysis of network data?  One popular solution comes from the class of exponential random graph models, or ERGMs \citep{frank_strauss_1986, wasserman_pattison_1996, snijders_2002, robins_pattison_kalish_etal_2007}.  These models assume a particular dependence regime for the full set of potential ties in the network.  In \citet{frank_strauss_1986}'s original formulation, Markov dependence is assumed (i.e., where all dyads that share a node are dependent), but more recent specifications have explored alternative dependence regimes \citep[e.g. local conditional Markov dependence in][]{snijders_pattison_robins_2006}.

Alternatively, we might consider representing this complex network data in an unobserved, latent lower-dimensional space and specifying a dependence regime there.  This approach draws on ideas developed by \citet{lazarsfeld_henry_1968} which proposed using latent variables to simplify complicated dependence structures and develop simpler model specifications: in these analyses, an unobserved latent variable is constructed such that the distribution of the data given the latent variables has a simple form. Of course, this idea has been used in a variety of settings, from factor analysis to item response theory to mixture models \citep{spearman_1904,vanderlinden_hambleton_2013,bartholomew_knott_moustaki_2011}, and so it is no surprise that such a technique might be helpful in the network data setting.  Additionally, we might consider choosing this lower dimensional space to be a metric space, in which case we are embedding the nodes of the network within a particular geometry \citep[e.g.][]{hoff_raftery_handcock_2002, mccormick_zheng_2015, airoldi_blei_fienberg_etal_2008, krioukov_papadopoulos_kitsak_etal_2010}.  Although the details of the embedding may vary across applications or analytical methods, nodes' positions in the latent space will correspond to features in the observed network with nodes that are connected in the network being closer together (or more generally, being more similar) in the latent space.  That is, dependence in the network can be induced through distances, or similarities, between nodes in this lower-dimensional latent space.  In practical settings, note that this framework provides a highly intuitive interpretation of the network data.  For example, in the social network setting, we can interpret the embedding as a map of nodes' relative positions in a latent ``social space,'' capturing the variety of unobserved social forces that may have influenced the observed pattern of social ties.

Traditionally, most latent space approaches have suggested embedding the nodes of the network in a Euclidean latent space \citep{hoff_raftery_handcock_2002}. Although Euclidean space is certainly familiar and relatively easy to work with, it is not clear that it is the most appropriate geometry for network data or that Euclidean space is best equipped to reveal the variety of potentially interesting network features \citep{krioukov_papadopoulos_kitsak_etal_2010}.  More specifically, we argue that the geometry of the latent space plays an important role in determining both the fitness of the analysis for the particular observed network data as well as the types of questions which the analysis is equipped to answer.  In order to tease out this relationship, we will consider relating observed network features to latent space geometry through two different lenses:  first, through common network summary statistics \citep[e.g., transitivity and average degree;][]{kolaczyk_csardi_2014} and second, through graph Laplacians which describe interesting features of networks and whose spectrum, under suitable assumptions of the latent space, is not influenced by network size.

We begin by summarizing existing approaches to the analysis of network data which take advantage of a lower-dimensional embedding of the nodes of the network in an unobserved latent space.  In Section 2, we discuss commonly used similarity measures, focusing on methodology that is most amenable to the traditional statistical modeling framework.  In Section 3.1, we review the variety of different types of latent spaces - from the familiar Euclidean space to more abstract spaces, like ultrametric space - in the continuous latent space model literature, in order to point out the impact the choice of geometry for the latent space can have on an analysis of network data.  We argue that the geometry of the latent space can play an important role in understanding network data, highlighting the advantages of a newly proposed latent space for network data, hyperbolic space, in Section 3.2.  In Section 4, we describe typical features of networks embedded in spaces with different curvatures using simulated network data.  In this way, we connect the geometry of the latent space to features of networks that we typically care about, such as centralization measures and properties of the degree distribution.  Finally in Section 5, we examine graph Laplacians to elucidate the relationship between the types of questions or discoveries that are possible in an analysis of network data which uses a latent space of a particular geometry.

%%%%%%%%%%%%%%%%%%%%%%%%%%%%%%%%%%
%%
%%    (  2  )   CLS MODELS FOR NETWORK DATA
%%
%%%%%%%%%%%%%%%%%%%%%%%%%%%%%%%%%%
\section{Continuous latent space models for network data} \label{secCLSsum}

A fully parametric model for network data provides a natural way to think about utilizing a lower-dimensional latent space for the nodes in the network.  A generic form for such a model is given below:

\begin{align} \label{eqLSMgen}
Y_{ij} &\overset{ind}{\sim} \text{Bernoulli}(p_{ij}) & i &\ne j; \>\> i,j = 1, ...,n \nonumber \\
\text{logit}(p_{ij}) & = \alpha + s(z_i,z_j)\\ 
z_i & \in \mathcal{Z}^t, \hspace{3mm} z_i  \overset{ind}{\sim}  f_t(z|\psi) &i &= 1,...,n, \nonumber
\end{align}
where $Y_{ij}$ is a binary indicator of a tie between nodes $i$ and $j$, $z_i$ is the $i$th node's $t$-dimensional unobserved latent vector in the $t$-dimensional space $\mathcal{Z}^t$, and $\psi$ parameterizes the distribution of the latent vectors in $\mathcal{Z}^t$.  The generic latent space model assumes that network ties are conditionally independent, given the dyad's latent vectors.  The probability of a tie consists of a baseline tendency for tie formation, $\alpha$, and depends on some similarity measure, $s(\cdot,\cdot)$, of the latent vectors so that the probability of a tie is higher for nodes with more similar latent vectors and lower for nodes with latent vectors that are dissimilar.  In this sense, the latent vectors can represent the observed pattern of ties in the network as well as any unobserved attributes of the nodes that may have contributed to the pattern of observed ties.  For example, in a social network, the latent vectors are typically interpreted as individuals' positions in a latent ``social space,'' representing the variety of unobserved social forces present in the observed network.  In most cases, the similarity measure, $s(\cdot,\cdot)$, is chosen so that transitivity is a natural consequence of the model specification; if nodes $i$ and $j$ are tied (i.e., $s(z_i,z_j)$ is large) and nodes $j$ and $k$ are tied (i.e., $s(z_j,z_k)$ is large), then $i$ and $k$ are likely tied (i.e., $s(z_i,z_k)$ is likely large).  This model can easily be generalized to incorporate homophily on observed attributes by simply adding a $\beta'x_{ij}$ term, where $x_{ij}$ is a vector of (dyad-level) observed covariates for the dyad or pair formed by the $i$th and $j$th nodes.  These latent space models typically require extra structure (e.g. hierarchical priors on the components of $\psi$) in order to model community structure and degree heterogeneity.

Here, we focus on methods that use a \textit{continuous} latent space (CLS), although the use of a discrete space is also possible \citep[e.g. the stochastic block models proposed by][described briefly in Section \ref{subsecDot}]{snijders_nowicki_1997}.  In practice, the dimension of the latent space, $t$, must be specified by the researcher, though traditional model selection techniques can be implemented to inform this choice.   Without loss of generality, we will assume $t=2$, which allows for easy visualization of the latent vectors and is a popular choice.  Note that this model can be easily extended to networks with non-binary (i.e. weighted) ties using ideas from generalized linear models.  Further, note that model \ref{eqLSMgen} is written for directed ties, and subsumes the setting of networks with undirected ties by simply replacing the condition $i \ne j$ with $i < j$ (though this may have implications for the choice of $s(\cdot,\cdot)$).  It is also worth pointing out that many existing network analysis methods which take advantage of a latent space representation do not require a fully parametric model, as outlined above, but instead fall under the umbrella of semi-parametric or non-parametric techniques.  We will highlight this distinction in our discussion of these methods below.

In the rest of this section, we describe existing latent space models for network data in terms of their similarity measure, $s(z_i,z_j)$.  In our discussion, we will group existing methods according to whether the proposed similarity measure has a dot product form, uses a distance metric, or is a graphon.  However, these categories are not strictly distinct.  For example, it may certainly be possible to rewrite a dot product model as a distance metric or perhaps to re-specify a distance model as a graphon.  Further understanding of the connections between these models provides an interesting avenue for future research (we will revisit this idea briefly in Section \ref{subsecGraph}).  Finally, we should point out that one of the most popularly cited papers regarding latent space models for networks, \citet{hoff_raftery_handcock_2002}, actually proposes two distinct versions of a model that uses a Euclidean latent space.  First, they propose a version which uses a distance metric (which they refer to as the ``distance model'' and which we will discuss later in Section \ref{subsecLit}) and then secondly, they propose a version which utilizes a dot product similarity measure (which they refer to as the ``projection model'' and which we discuss first, in the Section \ref{subsecDot}).

%%%%%%%%%%%%%%%%%%%%%%%%%%%%%%%%%%
	%%
	%%    (  2.1  )   DOT PRODUCT MODELS
	%%
%%%%%%%%%%%%%%%%%%%%%%%%%%%%%%%%%%
\subsection{Dot Product Models} \label{subsecDot}
\begin{figure}[t]
%trim = left bottom right top
  %\includegraphics[trim=0 19cm 0 0,width=.95\textwidth,clip]{epsVersions/DotPic.png}
   \includegraphics[trim=0 0cm 0 0,width=.95\textwidth,clip]{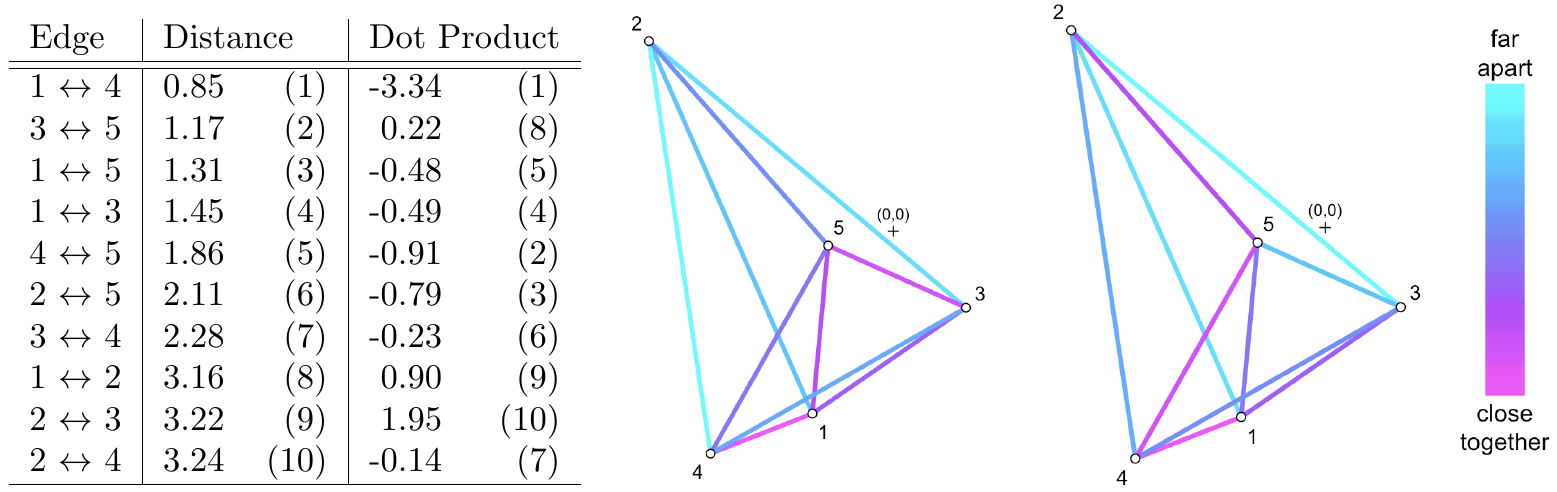}
  \caption[]{Euclidean Distance and the Dot Product.  For illustration, five points are simulated from a standard bivariate normal distribution.  In the network plots, edge color indicates the size of the similarity measure (Euclidean distance for the network on the left and dot product for the network on the right) between nodes, with magenta edges representing nodes that are least similar and cyan edges for those that are closest or most similar.  Note that the ranking of similarities across the two networks is not identical, for example consider the edge between nodes 3 and 5.}
  \label{figDot}
\end{figure}

Many latent space methods for network data use a simple dot product similarity measure,
\begin{align*}
s(z_i, z_j) &= z_i'z_j.
\end{align*}
In Euclidean space, the dot product measures the positioning of the vectors in the space.  For vectors in $\mathbb{R}^t$, the dot product is related to both the angular distance, $\phi$, and the Euclidean distance, $d(z_i,z_j)$, in the following ways:
\begin{align*}
z_i'z_j &= |z_i| |z_j| \text{cos} \phi \\
&= \frac{1}{2} \left\{ |z_i|^2 + |z_j|^2- \left[ d(z_i,z_j) \right]^2 \right\} .
\end{align*}
The dot product in $\mathbb{R}^t$ can be interpreted as the scaled cosine similarity (scaled by the magnitude of $z_i$ and $z_j$) or equivalently as a function of the squared distance between unit vectors in the direction of $z_i$ and $z_j$.  As can be seen in Figure \ref{figDot}, while the dot product in Euclidean space is certainly related to Euclidean distance, they are not equivalent measures.  Although Euclidean distance is certainly much easier to visualize, the dot product is simpler to calculate and thus has computational benefits.  

Most notably, the second version of the latent space model introduced in \citet{hoff_raftery_handcock_2002}, the so-called ``projection model'', fits within this framework (the first version of their proposed model uses a distance metric and will be discussed in detail in Section \ref{subsecLit}).  \citet{hoff_raftery_handcock_2002}'s projection model is specifically designed to handle directed network ties and is specified as follows:
\begin{align} \label{eqHoffProj}
s(z_i,z_j) &= - a_i z_i'z_j \\
z_i &\in \text{$t$-dimensional hypersphere}, \nonumber
\end{align}
where $a_i > 0$ is a sociality parameter for the $i$th individual.  Note that $s(z_i,z_j)$ can be interpreted as the signed scalar projection of $v_i$ onto $v_j$, where $v_i = a_i z_i$ and $v_j = a_j z_j$ are the scaled latent positions for the $i$th and $j$th node, scaled by their sociality parameters, $a_i$ and $a_j$ respectively \citep[see Figure \ref{figCircles}; note that the notation used here differs from that used in][]{hoff_raftery_handcock_2002}.  The sociality parameters model a node's tendency to send ties (i.e., to be ``social'').  Of course, the authors point out that this model could be easily adjusted to account for variability in nodes' tendency to receive ties, by replacing $a_i$ with $a_j$.

In practice, the authors suggest reparameterizing this model with alternative latent vectors, $v_i = a_i z_i$, so that
\begin{align*}
s(v_i,v_j) = \frac{v_i'v_j}{|v_j|} \\
v_i &\in \mathbb{R}^t.
\end{align*}
This specification of the model is more parsimonious since it does not require a separate model for the sociality parameters.  Further, since the $v_i$ are now arbitrary positions in $\mathbb{R}^t$, we can simply choose $f_t(v|\psi)$ to be a standard Normal distribution, as suggested for both versions of the latent space model proposed in \citet{hoff_raftery_handcock_2002}.

Other examples of dot product latent space models include \citet{mccormick_zheng_2015}'s model for aggregated relational data (ARD) on the hypersphere (discussed in more detail in Section \ref{subsecLit}) and \citet{hoff_2005}'s bilinear model, which incorporates additive node-level sociality effects to allow for degree heterogeneity,
\begin{align*}
s(z_i, z_j) &= a_i + b_j + z_i'z_j \\
a_i &\overset{iid}{\sim}\text{Normal}(0,\sigma^2_a), \>\> b_i \overset{iid}{\sim}\text{Normal}(0,\sigma^2_b)\\
z_i &\overset{iid}{\sim}\text{Multivariate Normal}(0,\sigma^2 I_t),
\end{align*}
where $a_i$ and $b_j$ are sociality effects for the $i$th and $j$th node respectively.  \citet{hoff_2005} motivates the interpretation of $\sigma^2_a$, $\sigma^2_b$, and $\sigma^2$ as random effects variances, highlighting similarities between the model and a traditional ANOVA model.  In fact, these sociality effects mimic the similarity measures used in later models \citep[e.g.][see Section \ref{subsecLit} for more discussion]{krivitsky_handcock_raftery_hoff_2009}, but the idea of modeling sender and receiver effects has a long history \citep{holland_leinhardt_1981, fienberg_wasserman_1981, vanduijn_snijders_zijlstra_2004}. 

Dot product models in Euclidean space have been further developed by \citet{young_scheinerman_2007}, \citet{nickel_2007}, and \citet{pao_coppersmith_priebe_2011} as the class of random dot product graphs, where $\mathcal{Z}^t$ is the $[0,1]$ box in $\mathbb{R}^t$ and $s(z_i,z_j)$ is the dot product, or some function of it.

We can also further generalize the type of similarity measure used in dot product models to a quadratic-like form,
\begin{align}
s(z_i,z_j) = z_i'Az_j,  \label{eqSimQuad}
\end{align}
where $A$ is a $t \times t$-dimensional matrix.  This specification subsumes the models discussed previously in this section, by simply choosing $A$ to be the identity matrix.  Examples include  \citet{hoff_2008}'s eigenmodel for undirected networks,  \citet{hoff_2009}'s SVD-inspired extension to directed networks, and \citet{minhas_hoff_ward_2016}'s class of additive and multiplicative effects (AME) models.

It is also worth pointing out that the traditional stochastic block model (SBM) can be viewed as a special case of these quadratic-like latent space models.  Generally, the stochastic block model assumes that the nodes can be separated into $t$ unobserved latent classes where the membership vectors, $z_i$, are unknown.  From model \ref{eqLSMgen}, assume $\alpha=0$ and let
\begin{align*}
s(z_i,z_j) &= p_{\pi(z_i),\pi(z_j)}\\
z_i &\sim \text{Multinomial}_t(1,\psi),
\end{align*}
where $\pi$ maps nodes' latent vectors to $t$ classes, $p_{t_1 t_2}$ parameterizes ties between nodes in the $t_1$th and $t_2$th classes, and $\psi$ is a $t$-dimensional vector of membership probabilities.  The model treats nodes within the same class as stochastically equivalent and parameterizes the probability of ties between nodes belonging to different classes by the $p_{t_1 t_2}$s.  Note that we can rewrite the similarity function in the quadratic-like form of equation \ref{eqSimQuad} by letting $A$ be a $t \times t$ matrix of the $\left\{ p_{t_1 t_2} \right\}$ parameters.  In the original versions of this model, the number of classes $t$ was assumed to be known, though a variety of methods now exist to help understand the implication of different choices of $t$ \citep{saldana_yu_feng_2015,chen_lei_2016}.

%If the vertices are (re)ordered by their memberships, then $A$ is a block matrix given by
%\begin{align*}
%\left[ \begin{array}{cccc}
%p_{11} J_{n_1} & p_{12} J_{n_1, n_2} & \cdots & p_{1t} J_{n_1, n_t} \\
%p_{21} J_{n_2, n_1} & p_{22} J_{n_2} & \cdots & p_{2t} J_{n_2, n_t} \\
%\vdots & \vdots & \ddots & \vdots \\
%p_{t1} J_{n_t, n_1} & p_{t2} J_{n_t, n_2} & \cdots & p_{tt} J_{n_t} \\
%\end{array} \right],
%\end{align*}
%where $J$ is a matrix of ones and $n_t$ is the size or number of members in class $t$. 

\citet{airoldi_blei_fienberg_etal_2008} further extend this model by allowing nodes to have simultaneous partial memberships in each of the $t$ classes in the mixed membership stochastic block model (MMSBM) with
\begin{align*}
s(z_i,z_j) &= z_i'Az_j\\
z_i &\sim \text{Dirichlet}_t(\psi),
\end{align*}
where again $A$ is the block diagonal matrix of the between-class tie probability vector and $\psi$ is a $t$-dimensional vector of membership probabilities.  In this setting, the $z_i$s are still interpreted as unobserved latent membership vectors and further, can be viewed geometrically as positions on a $(t-1)$-dimensional simplex.  In fact, in the original specification of the stochastic block model, the $z_i$ can be viewed geometrically as the \textit{vertices} of a $(t-1)$-dimensional simplex.  In the MMSBM, the membership vectors have a geometric interpretation on the simplex:  distance from a node's position on the simplex to a particular vertex (community) is an indication of how strongly that node's behavior in the network matches the behavior of the community as a whole.

%%%%%%%%%%%%%%%%%%%%%%%%%%%%%%%%%%
	%%
	%%    (  2.2  )   DISTANCE MODELS
	%%
%%%%%%%%%%%%%%%%%%%%%%%%%%%%%%%%%%
\subsection{Distance Models} \label{subsecDist}

A natural choice for $s(\cdot,\cdot)$ is a (negative) distance metric, in which case we can interpret the $z_i$s as latent positions in a $t$-dimensional latent metric space, $\mathcal{Z}^t$.  We will refer to these special cases as latent distance models:
\begin{align} \label{eqLDMgen}
Y_{ij} &\overset{ind}{\sim} \text{Bernoulli}(p_{ij}) & i &\ne j; \>\> i,j = 1, ...,n, \nonumber \\
\text{logit}(p_{ij}) & = \alpha - d(z_i,z_j) \\
z_i & \in \mathcal{Z}^t, \hspace{3mm} z_i  \overset{ind}{\sim}  f_t(z|\psi) &i &= 1,...,n, \nonumber
\end{align}
where $d$ is the natural metric on $\mathcal{Z}^t$ and $f_t$ describes the distribution of latent positions in $\mathcal{Z}^t$.  In a latent distance model, transitivity effects are guaranteed by the triangle inequality.  Note also that these models are most appropriate for undirected ties, since distances are inherently symmetric.  These models also allow us to visualize the nodes' latent positions by simply plotting estimates, $\hat{z_i}$, in $\mathcal{Z}^t$ which can help illuminate previously unnoticed patterns in the data.  For example, one could plot the latent positions from a null model and color the positions according to some covariate that could later be included in the model.  Although this procedure is available to all versions of latent space models (whether distance is the chosen similarity measure or not), interpretation of relationships among nodes' latent positions in these plots is certainly more intuitive in the latent distance models.

Further, note that the concept of using distance to model dependence is not unique to the class of latent distance models for network data.  In fact, this idea has had a long tradition of success in the statistical modeling literature.  For example, in methods for continuously-indexed spatial data, distance is obviously a natural part of the modeling framework.  Even in cases where spatial data is observed at an aggregate level (i.e., in some type of districts or areal units) the spatial weights matrix used in spatial regression models, such as CAR and SAR models, contains information about distance through a nearest neighbors matrix (where the $ij$th entry is one if the $i$th district neighbors the $j$th district and zero otherwise).  In fact, in many spatial applications, non-Euclidean distances can be used to better represent the underlying dependence structure of the particular data generating process.  For example, a researcher might use stream or flow distance when considering chemical concentrations in a watershed area or perhaps travel time or travel cost for a group of city commuters. Further, models for phylogenetic data often make use of Hamming distance\footnote{The Hamming distance between two vectors counts the number of positions at which the corresponding vectors differ.} to represent similarity among species when creating a phylogenetic tree.  In short, using latent distances to model dependence among nodes in a network allows for an intuitive representation of an otherwise complex dependence structure and draws on the well-developed practices of using distance to model dependent data in other fields of statistical research.

Examples of latent space approaches which use distance include \citet{hoff_raftery_handcock_2002, handcock_raftery_tantrum_2007, krivitsky_handcock_raftery_hoff_2009} and will be discussed in more detail in Section \ref{subsecLit}.

%%%%%%%%%%%%%%%%%%%%%%%%%%%%%%%%%%
	%%
	%%    (  2.3  )   GRAPHONS
	%%
%%%%%%%%%%%%%%%%%%%%%%%%%%%%%%%%%%
\subsection{Graphons} \label{subsecGraph}
%Graphons are essentially equivalent to CLS models.
%To be precise, a \textit{graphon} is essentially a measurable function $W:[0,1]^2\rightarrow[0,1]$ symmetric in its coordinates \cite{lovasz2012large}.
%A graphon can be regarded as the CLS model given in \ref{eqLSMgen} by taking the $z_i$'s to be uniformly and independently drawn from $[0,1]$ and setting $p_{ij}=W(z_i,z_j)$.
% A CLS model, at least in the oft-case where $[0,1]$ can be identified with the latent space $\mathcal{Z}^t$ along a (likely discontinuous) isomorphism mod $0$ of probability measure spaces, can be regarded as a graphon by setting $W(z_i,z_j)=p_{ij}$.
%Mixtures of graphons are exactly the distributions of infinite graphs invariant under permutations of the nodes \cite{lovasz2012large}. 
%Conversely, it is possible to formalize a sense in which a limit of suitable sequences of growing graphs uniquely identifies a graphon up to graphon equivalence \cite{diaconis2007graph}.
%The use of graphons as models of growing networks is broad and growing.

Graphons provide a different perspective than the models discussed thus far, but in many cases, the graphon approach is essentially equivalent to the CLS approach.
To be precise, a \textit{graphon} is a measurable function $W$ which maps $[0,1]^2$ to $[0,1]$, and is symmetric in its coordinates \citep{lovasz2012large}.  The graphon approach can be regarded as the CLS model given in \ref{eqLSMgen} by taking the $z_i$'s to be uniformly and independently drawn points in the unit interval and setting $p_{ij}=W(z_i,z_j)$.
 On the other hand, a CLS model can be regarded as a graphon by setting $W(z_i,z_j)=p_{ij}$, as long as (and is often the case that) the unit interval can be identified with the CLS model's latent space, $\mathcal{Z}^t$, along a (likely discontinuous) isomorphism mod $0$ of probability measure spaces.
In this sense, we can think of the graphon approach as a CLS model with a simple latent space -- the unit interval -- but with a very general class of (often complicated) functions, $W$, that specify the connection probabilities.  
Graphons are commonly used to study asymptotic behavior and networks that grow in size.  
For example, mixtures of graphons are exactly the distributions of infinite graphs invariant under permutations of the nodes \citep{lovasz2012large}. 
Conversely, it is possible to formalize a sense in which a limit of suitable sequences of growing graphs uniquely identifies a graphon up to graphon equivalence \citep[i.e., up to relabeling;][] {diaconis2007graph}.

%However, graphons and CLS models often differ in how they are typically grouped together in families from which network inference is conducted.
%When inferring a CLS model, the allowable CLS models to consider typically share a fixed latent space, fixed connection probabilities, and differ only in the node densities \citep[c.f.][]{hoff_raftery_handcock_2002, krioukov_papadopoulos_kitsak_etal_2010}.  When inferring a graphon, the allowable graphons necessarily differ only in their connection probabilities.   In order to increase the efficiency of the estimation process, it is desirable to put constraints on the allowable models.

While the specification of graphon and CLS models are similar, the standard inferential strategies for the two types of models vary.  Inference in CLS models focuses on the estimation of unknown parameters characterizing the distribution of the positions in latent space, in addition to parameters capturing standard fixed and random effects components of the model.  On the other hand, inference for graphon models focuses on estimation of the $W$ function; in CLS models, the functional form of the connection probabilities is treated as fixed.  In many settings, constraints on allowable graphon models are used to increase the efficiency of the estimation process.  For example, a graphon can be practically estimated from samples of a stochastic blockmodel \citep{abbe_bandeira_hall_2016,amini2013pseudo,bickel2009nonparametric}, efficiently under an assumption of H\"{o}lder continuity on the $W$-function \citep{gao2015rate,wolfe2013nonparametric}.
However, it is difficult to impose a simple condition on a $W$-function, in order to constrain salient network characteristics of interest.
In contrast, choice of latent space and constraints on the the parameters of the distribution of latent positions in a CLS model can impose natural constraints on small world properties, degree distributions, and clusterability \citep[c.f.][]{krioukov_papadopoulos_kitsak_etal_2010}.

Graphons and CLS  models also differ in how they are used to compare random networks. Two graphons are often compared by a \textit{cut-metric}, an infima of $L_p$-distances between the $W$-functions \textit{up to graphon equivalence} \citep{lovasz2012large}.
Two CLS distance models with a manifold as a fixed latent space and connection probability are often compared by an infima of $L_p$-distances between the node densities \textit{up to a distance-preserving bijection} of latent space.
The latter distance implicitly takes into account the geometry of the latent space; the former distance, only dependent up to measure-preserving bijections, cannot possibly take into account any interesting features of the latent space.

%%%%%%%%%%%%%%%%%%%%%%%%%%%%%%%%%%
%%
%%    (  3  )   CLS MODELS AND LATENT SPACE GEOMETRY
%%
%%%%%%%%%%%%%%%%%%%%%%%%%%%%%%%%%%
\section{Latent space geometries} \label{secCLSgeom}

In this section, we continue our description of continuous latent space models for network data, paying particular attention to the geometric properties of the latent space assumed under each model.  In Section \ref{subsecLit}, we review existing approaches in Euclidean space (Section \ref{subsubsecEuclid}); elliptic space, such as the surface of a three-dimensional sphere (Section \ref{subsubsecEll}); and other spaces (Section \ref{subsubsecOther}).  In Section \ref{subsecHyper}, we highlight properties of hyperbolic space and detail its promising potential application in the class of continuous latent space models for network data, inspired by recent network analysis methods which take advantage of a hyperbolic latent space \citep{krioukov_papadopoulos_kitsak_etal_2010,asta_shalizi_2014,aldecoa_orsini_krioukov_2015}.  We will discuss various interesting aspects of the different choices for the geometry of the latent space in each subsection, but we also have provided a summary of some of the most important differences between Euclidean, elliptic, and hyperbolic space in Table \ref{tabCurve}.

%%%%%%%%%%%%%%%%%%%%%%%%%%%%%%%%%%
	%%
	%%    (  3.1  )   LITERATURE REVIEW
	%%
%%%%%%%%%%%%%%%%%%%%%%%%%%%%%%%%%%
\subsection{Literature Review} \label{subsecLit}

%%%%%%%%%%%%%%%%%%%%%%%%%%%%%%%%%%
		%%
		%%    (  3.1.1  )   EUCLIDEAN SPACE
		%%
%%%%%%%%%%%%%%%%%%%%%%%%%%%%%%%%%%
\subsubsection{Euclidean Space} \label{subsubsecEuclid}
Perhaps the most intuitive specification of the latent space model is the Euclidean latent distance model, where $\mathcal{Z}^t$ is the familiar $t$-dimensional Euclidean space and $d$ is Euclidean distance,
\begin{align*}
d(z_i,z_j) &= |z_i - z_j|\\
& = \sqrt{ (z_i-z_j)'(z_i-z_j) }.
\end{align*}
In fact, this version of the model was introduced by \citet{hoff_raftery_handcock_2002} as the ``distance model'' and is certainly the most widely used latent space model for network data.  The authors specify $f_t(z|\psi)$ as a standard normal distribution. 

\citet{handcock_raftery_tantrum_2007} extend this model to capture community structure in a network by letting $f_t(z|\psi)$ be a mixture of $K$ multivariate normal distributions:
\[ z_i \sim \sum_{k=1}^K \lambda_k \text{MVN}_t \left( \mu_k, \sigma^2_k I_t \right), \]
where $\lambda_k$ is the probability that a node belongs to the $k$th group ($\sum_{k=1}^K \lambda_k =1$).  Under this model, the latent positions are spatially clustered in $\mathcal{Z}^t$.  Thus, this model can also be viewed as a generalization of the stochastic block model (see Section \ref{subsubsecOther} for more details), with the $k$th mixture component representing the $k$th class, which also allows for transitivity within blocks and homophily on attributes.  On the other hand, \citet{handcock_raftery_tantrum_2007}'s model can be thought of as an extension of \citet{hoff_raftery_handcock_2002}'s version of the Euclidean latent distance model which introduces more structure in the latent positions (e.g. spatial clustering) to better represent potential community structure in the observed network.  

Although developed for directed ties, \citet{handcock_raftery_tantrum_2007}'s model does not explicitly model degree heterogeneity.  This extension is provided by \citet{krivitsky_handcock_raftery_hoff_2009} where the addition of additive individual random effects act as sociality or popularity effects to accommodate the differing tendency of nodes to send or receive ties, with
\begin{align*}
s(z_i,z_j) &= a_i + b_j - d(z_i,z_j)\\
a_i & \overset{iid}{\sim} \text{Normal}(0,\sigma^2_a), \>\>\>\> b_i \overset{iid}{\sim} \text{Normal}(0,\sigma^2_b),
\end{align*}
where $a_i$ is the sender effect for the $i$th node and $b_j$ is the receiver effect for the $j$th node.  The authors' extension of \citet{handcock_raftery_tantrum_2007}'s model borrows from the form of \citet{hoff_2005}'s bilinear model, mentioned in Section \ref{subsecDot}.  In both cases, the sender and receiver effects mimic the specification of an ANOVA model and can be interpreted similarly.

The latent Euclidean distance model has also been extended to the case of dynamic networks \citep{sarkar_moore_2006, sewell_chen_2015}.  % Need a brief description and model statement (maybe?)

While a Euclidean latent space certainly allows for easy visualization and interpretation of estimated latent positions and may be desirable given the familiarity of Euclidean geometry, it is not clear whether this geometry is best suited to represent the complex dependencies we typically encounter in network data.  We will discuss this point further in Section \ref{subsecHyper}.

%%%%%%%%%%%%%%%%%%%%%%%%%%%%%%%%%%
		%%
		%%    (  3.1.2  )   ELLIPTIC SPACE
		%%
%%%%%%%%%%%%%%%%%%%%%%%%%%%%%%%%%%
\subsubsection{The Unit Hypersphere} \label{subsubsecEll}

As discussed in Section \ref{subsecDot}, \citet{hoff_raftery_handcock_2002} introduce two versions of a latent space model:  the distance model in Euclidean space (see Section \ref{subsubsecEuclid}) and the so-called projection model, which uses a dot product similarity measure and embeds nodes on the unit circle (i.e., a 1-dimensional unit hypersphere).

\begin{figure}[t]
\includegraphics[width=.95\textwidth]{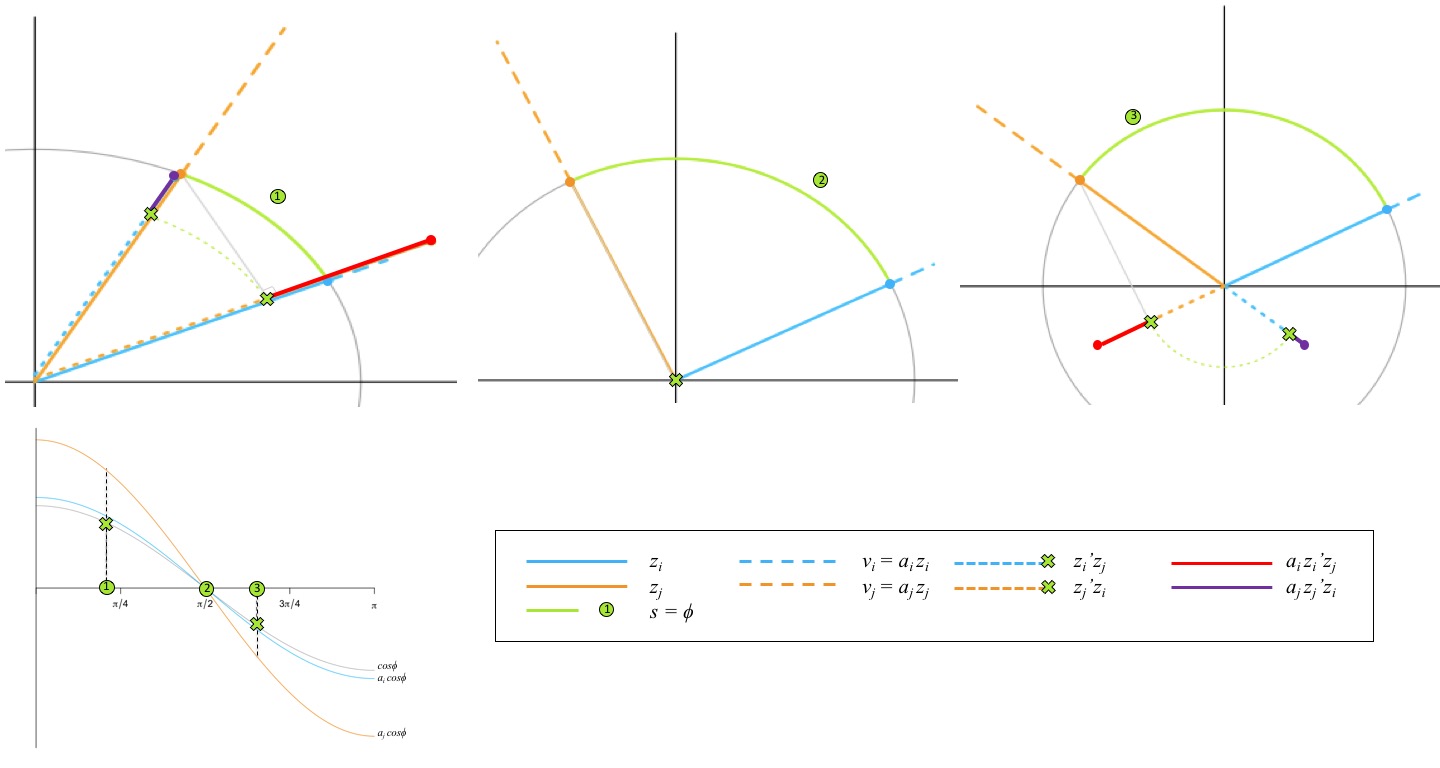}
%\vspace{-.25in}
\caption{Visual interpretation of the choice of $s(\cdot,\cdot)$ in \citet{hoff_raftery_handcock_2002}'s projection model, $t=2$.  Latent position vectors for the $i$th node are shown in blue and for the $j$th node are shown in orange.  Across the top panel, we display three possible sets of latent positions, with their corresponding similarity measure in \citet{hoff_raftery_handcock_2002}'s model identified by green crosses.  Distance in this latent space, the unit hypersphere, is great circle distance or arc length and is displayed as green arcs in the top panels and green points in the bottom left panel.}\label{figCircles}
\end{figure}

Note that this projection model is \textit{not} an example of a latent distance model, since the natural distance metric on the hypersphere is great-circle distance (arc length).  However, on the unit hypersphere, the great-circle distance between two latent positions, $z_i$ and $z_j$, is equivalent to their angular distance, $\phi$, and can be calculated as the inverse cosine of the dot product of $z_i$ and $z_j$.  Thus, the similarity measure in \citet{hoff_raftery_handcock_2002}'s projection model (see Equation \ref{eqHoffProj}) is simply the \textit{cosine similarity} between the latent position vectors, scaled by a sociality parameter, $a_i$ (See Figure \ref{figCircles} for some examples).  When $z_i$ and $z_j$ are close (and the angle between them is acute, as in the left panel of Figure \ref{figCircles}), $z_i'z_j$ will be positive and when they are far apart (and the angle between them is obtuse, as in the right panel of Figure \ref{figCircles}), $z_i'z_j$ will be negative.  Thus, this choice of $s(\cdot,\cdot)$ preserves the relationship between the probability of a tie and some measure of ``closeness.''  That is, just as in the latent distance models, nodes that are ``closer'' together are more likely to be tied.  Further, if this model were used for networks with undirected ties, the sociality parameters (the $a_i$s) would disappear, leaving $s(\cdot,\cdot)$ as the cosine similarity, which is a monotonic function of the natural distance metric, $\phi$, on the unit hypersphere.  However, although cosine similarity is not a proper distance metric itself (since, for example, it is not nonnegative), it is a popular similarity measure.  Compared to a true latent distance model on the unit hypersphere, this choice of $s(\cdot,\cdot)$ changes the interpretation of $\alpha$ (and any parameters for covariates, if included) in model \ref{eqLSMgen} and also re-weights the distances, since cos$\phi$ is a non-linear function of $\phi$.  For example, given a fixed network, we would not expect an estimate of $\alpha$ under the distance model to match an estimate under the projection model.  Generically, $\alpha$ represents a baseline tendency for tie formation, \textit{relative to the average expected similarity} under the chosen model.  For example, consider a two-dimensional latent space and assume that each dimension of the latent positions are independently and identically distributed according to a standard Normal, it is relatively easy to show that under the distance model the average expected similarity is $\sqrt{\pi}$ while under the projection model this same quantity is zero.

\citet{mccormick_zheng_2015} extend the projection model to the setting of aggregated relational data (ARD), an example of incomplete network data.  In this setting, surveyed individuals answer ``how many $X$'s do you know?,'' where $X$ represents particular, typically hard-to-reach, groups of interest such as drug users, terrorists, or homeless individuals.  Thus, rather than observe the entire network, we only observe connections between surveyed individuals and the groups of interest;  we do not observe ties between surveyed individuals and actual members of the group nor ties between members of the group.  To accommodate this type of data, the authors first propose a model for the complete case (if the network had been entirely observed) and then derive a model for the aggregated observations.  The proposed model for the complete case is a slight extension of \citet{hoff_raftery_handcock_2002}'s projection model, where cosine similarity is again incorporated:
\[ s(z_i,z_j) = a_i + a_j + \zeta z_i'z_j, \]
where $a_i$ is a gregariousness or sociality effect for the $i$th node and $\zeta$ scales the overall influence of the latent component.  Compared to \citet{hoff_raftery_handcock_2002}'s projection model, \citet{mccormick_zheng_2015} specify a sociality effect that is additive rather than multiplicative and that accounts for differences between the sender and the receiver.  They also incorporate an additional scale parameter, $\zeta$, which modifies the influence of the latent vectors.  Latent positions for the survey participants are modeled as uniformly distributed across the hypersphere.

%%%%%%%%%%%%%%%%%%%%%%%%%%%%%%%%%%
		%%
		%%    (  3.1.3  )   OTHER SPACES
		%%
%%%%%%%%%%%%%%%%%%%%%%%%%%%%%%%%%%
\subsubsection{Other Spaces} \label{subsubsecOther}

Although both the SBM and MMSBM allow for geometric interpretations of the latent vectors (i.e., as positions on the simplex), neither is an example of a latent distance model\footnote{Of course, utilizing $s(z_i,z_j) = -d(z_i,z_j)$ in the original stochastic block model is meaningless, since all vertices of the simplex are, by design, equidistant.  In the mixed-membership stochastic block model, this choice would significantly change the interpretation of the model:  from a node's tie probabilities depending on a weighted sum of the between-class tie probabilities, weighted by the node's membership vector, to a node's tie probabilities depending on distances in the space of all membership vectors.}.  Although some metrics have been developed in the context of compositional data, they generally are only intended for points on the open simplex and exclude points on the boundary.  For example, of the popular Aitchison distance metric defined for the open simplex, \citet{aitchison_barcelo-vidal_martin-fernandez_etal_2000} note that when one of the coordinates of a point in the simplex tends to zero, the distance between it and other points in the simplex will tend toward infinity.  The authors explain, ``There is nothing surprising about this feature; it is merely recognizing that a composition with one of the parts absent may be chemically, physically, or biologically completely different from compositions with all components positive. Doveton's (1998) perfect martini with its (gin, dry martini, sweet martini) composition is completely different from a cocktail with no gin but only dry and sweet martini present.''  For compositional data, this may be a natural assumption but it is not clear whether it is necessary in the setting of latent membership vectors.

In \citet{schweinberger_snijders_2003}, the authors specify a latent distance model in an ultrametric latent space.  Unlike other versions of the latent space models considered here, this model does not specify $\mathcal{Z}^t$ as a traditional metric space.  Recall that a metric space consists of a set, $M$, and distance (metric), $d$, which satisfies the following properties for all points $x,y,$ and $z \in M$:
\begin{enumerate}
\item Non-Negativity:  $d(x,y) \ge 0$.
\item Identity of Indiscernibles:  $d(x,y)=0 \iff x=y$.
\item Symmetry: $d(x,y) = d(y,x)$.
\item Triangle Inequality:  $d(x,z) \le d(x,y) + d(y,z)$. 
\end{enumerate}

\begin{figure}[t]
%trim = left bottom right top
\includegraphics[width=1\textwidth,trim=0in 4.75in 0in 4.25in,clip]{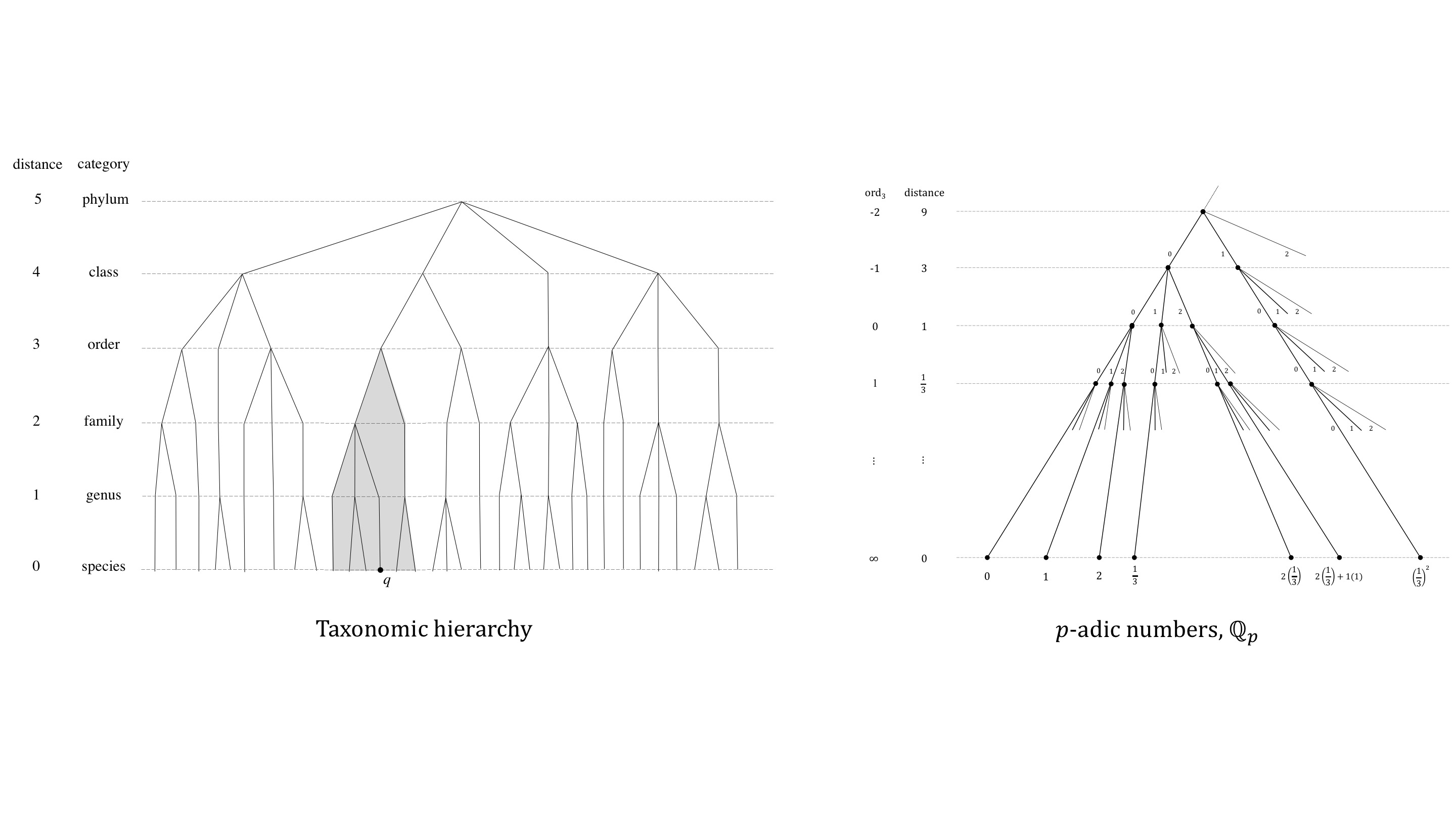}
\caption{Examples of Ultrametric Space.  In the left panel, we illustrate how the animal species classification system can be viewed as an ultrametric space, highlighting a disc of radius $\gamma=3$ centered at species $q$.  In the right panel, we display $\mathbb{Q}_p$, the field of $p$-adic numbers, taking $p=3$.  Roughly speaking, two points in $\mathbb{Q}_p$ are ``close" if their difference is divisible by a large positive power of $p$.  In the panel above, for example, 0 and $1$ are closer together than are 0 and $\frac{1}{3}$, because $1-0=1=3^0$ while $\frac{1}{3}-0= 3^{-1}$.  When visualized as a dendrogram, distance in ultrametric spaces is represented by the height of the smallest (inverted) tree between two members (members are represented as points on the bottom-most line in each panel).}
\label{figTree}
\end{figure}

An ultrametric space, or non-Archimedean space, is a metric space where $d$ additionally satisfies the strong triangle inequality:
\[ d(x,z) \le \text{max} \left\{ d(x,y), d(y,z) \right\}, \]
which, among other things, means that all triangles are isosceles with the unequal side being shortest and that every point in a disc is a center of that disc, so that two discs intersect only if one disc completely contains the other.  Not surprisingly, visualizing an ultrametric space and understanding the behavior of distances within it can be difficult.  However, ultrametric space can be visualized via dendrograms or trees.  For example, $\mathbb{Q}_p$, the field of $p$-adic numbers, is a well-known ultrametric space which can be visualized as the tree in the right panel of Figure \ref{figTree} \citep[see][for a thorough motivation of this figure]{holly_2001}.  For simplicity, consider a tree of the familiar animal species classification system in the left panel of Figure \ref{figTree}.  Naturally, the distance between species is the height of the smallest (inverted) tree between them so that, for example, species within the same genus are closer (d=1) than those who only share a common order (d=3).  Returning to the intersection of discs in ultrametric space, we illustrate one example of a disc visualized in the ultrametric tree for the animal species classification system in the left panel of Figure \ref{figTree}.  For species $q$, the disc centered at this species with radius $\gamma=3$ is the set of all leaves in the subtree descending from the node at distance level $\gamma$.  If we consider any arbitrary point (or species) within this disc, say $x$, then the disc centered at $x$ is the set of leaves in the subtree descending from the (unique) node above $x$ at distance level $\gamma$; this yields the exact same disc.

\begin{figure}[t]
%trim = left bottom right top
\includegraphics[width=1\textwidth,page=2,trim=0.65in 2.25in 0.65in 1.75in, clip]{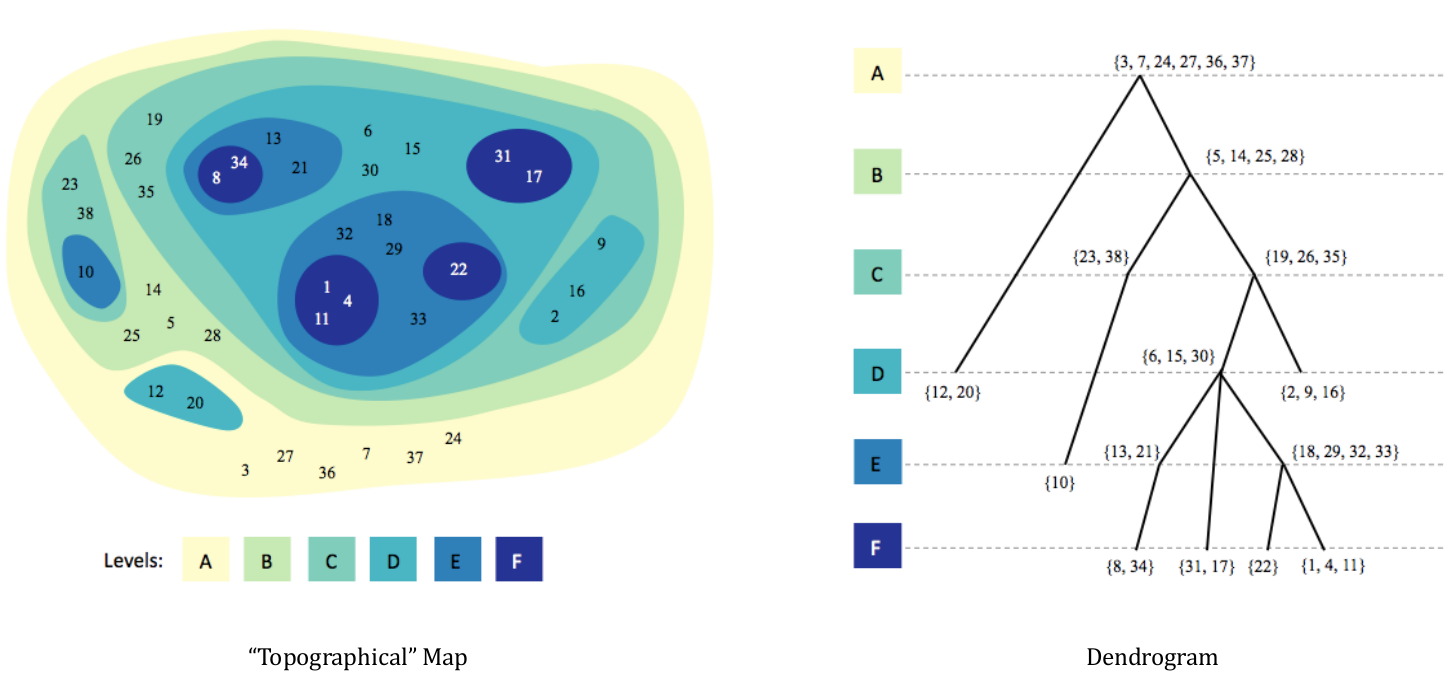}
\caption{Positions in a Latent Ultrametric Space.  As a toy example, we plot ``positions'' in a latent ultrametric space, visualized above in the style of a topographical map, on the left, and as a dendrogram, on the right.  Colors represent the levels and the nodes are indexed arbitrarily.  Note that nodes' relative positions within a particular level in the topographical map are meaningless and are simply an artifact of this particular visualization.}
\label{figUM}
\end{figure}

\citet{schweinberger_snijders_2003} motivate this choice for $\mathcal{Z}^t$ by noting that social scientists' understanding of latent settings\footnote{Here, the authors refer to the settings terminology emphasized by \citet{pattison_robins_2002} where a setting is a close-knit cluster of actors that are strongly tied.} structures mimic properties guaranteed in ultrametric spaces.  More specifically, the latent settings should be non-overlapping,  the interaction within settings should be stronger than the interaction between settings, and hierarchical nesting of the settings is expected.  The authors fit the model to a fraternity interaction network and provide a topological map (e.g. Figure \ref{figUM}) of the estimated latent positions where each level of the mapped social ``mountain'' can be interpreted as a social setting.  An example of the style of resulting settings groups is provided in Figure \ref{figUM}

%%%%%%%%%%%%%%%%%%%%%%%%%%%%%%%%%%
	%%
	%%    (  3.2  )   CLS MODELS IN H2
	%%
%%%%%%%%%%%%%%%%%%%%%%%%%%%%%%%%%%
\subsection{Hyperbolic Space} \label{subsecHyper}

Recent work in latent space models for network data has resulted in a novel proposal:  the use of negatively curved $t$-dimensional hyperbolic space, $\mathbb{H}^t$.  In the following section, we review some recent work in this area and provide some intuition for why hyperbolic space may be particularly amenable to modeling important network characteristics.  We refer the reader to Appendix \ref{HypModels.sec} for more details about the general geometric properties of hyperbolic space.  First suggested by \citet{krioukov_papadopoulos_kitsak_etal_2010}, modeling and analyzing networks in latent hyperbolic spaces is rapidly being investigated \citep[][among others]{aldecoa_orsini_krioukov_2015,asta_shalizi_2014,smith_2017}.  A generic latent distance model in $2$-dimensional hyperbolic space can be written as follows:
\begin{align}
Y_{ij} &\overset{ind}{\sim} \text{Bernoulli}(p_{ij})  \nonumber \\
\text{logit}(p_{ij}) & = \alpha - d(z_i,z_j) \label{hypModel.eq} \\ 
z_i & \in \mathbb{H}^2, \hspace{3mm} z_i  \overset{ind}{\sim}  f_2(z|\psi),  \nonumber %\\[.25in]
%\text{where} \hspace{5mm} d(z_i,z_j) &= d \left( (r_i,\phi_i), (r_j,\phi_j) \right) \nonumber \\
%&= \text{acosh} \left\{ \text{cosh}(r_i) \text{cosh}(r_j) - \text{sinh}(r_i) \text{sinh}(r_j) \text{cos}\left( \Delta \phi \right) \right\} \label{hypDist.eq} \\%[.25in]
%\text{and} \hspace{7mm} \Delta \phi &= \pi - | \pi - |\phi_i - \phi_j ||, \nonumber
%\end{align}
\end{align}
where
\begin{align}
d(z_i,z_j) &= d \left( (r_i,\phi_i), (r_j,\phi_j) \right) \nonumber \\
&= \text{acosh} \left\{ \text{cosh}(r_i) \text{cosh}(r_j) - \text{sinh}(r_i) \text{sinh}(r_j) \text{cos}\left( \Delta \phi \right) \right\}, \label{hypDist.eq} %\\%[.25in]
\end{align}
and
\begin{align*}
\Delta \phi &= \pi - | \pi - |\phi_i - \phi_j ||, \nonumber
\end{align*}
so that $d(z_i, z_j)$ is the appropriate distance metric in hyperbolic space (using polar coordinates for convenience) and $f_2(z|\psi)$ is a distribution for positions in $2$-dimensional hyperbolic latent space (extending this specification to $t$-dimensional hyperbolic space is trivial).  Of course, the exact form of this model and the choice of $f_t(z|\psi)$ can be altered - perhaps to respect particular network features or to accommodate different network modeling assumptions.

\citet{krioukov_papadopoulos_kitsak_etal_2010} provides a thorough motivation for the use of hyperbolic space in the class of latent space models for network data, stemming from the observation that many real world networks tend to be tree-like and that the geometry of a tree behaves much like the geometry of hyperbolic space. Informally, trees can be thought of as discrete hyperbolic spaces since $\mathbb{H}^2$ can be constructed as metrically equivalent to $b$-ary trees, trees with a branching factor of $b$. Further, the number of nodes at distance (path length) $r$ in a $b$-ary tree grows exponentially. Thus, trees need an exponential amount of space for branching and only hyperbolic space is able to accommodate this since hyperbolic space is ``bigger'' than Euclidean space.  For example, consider linearly increasing the radius of a sphere;  in Euclidean space, the circumference of the sphere also grows linearly, but in hyperbolic space the circumference grows exponentially (see Table \ref{tabCurve}). 

\begin{figure}[t]
%trim = left bottom right top
\includegraphics[width=1\textwidth,page=3,trim=0.65in 1.75in 0.65in 1.5in, clip]{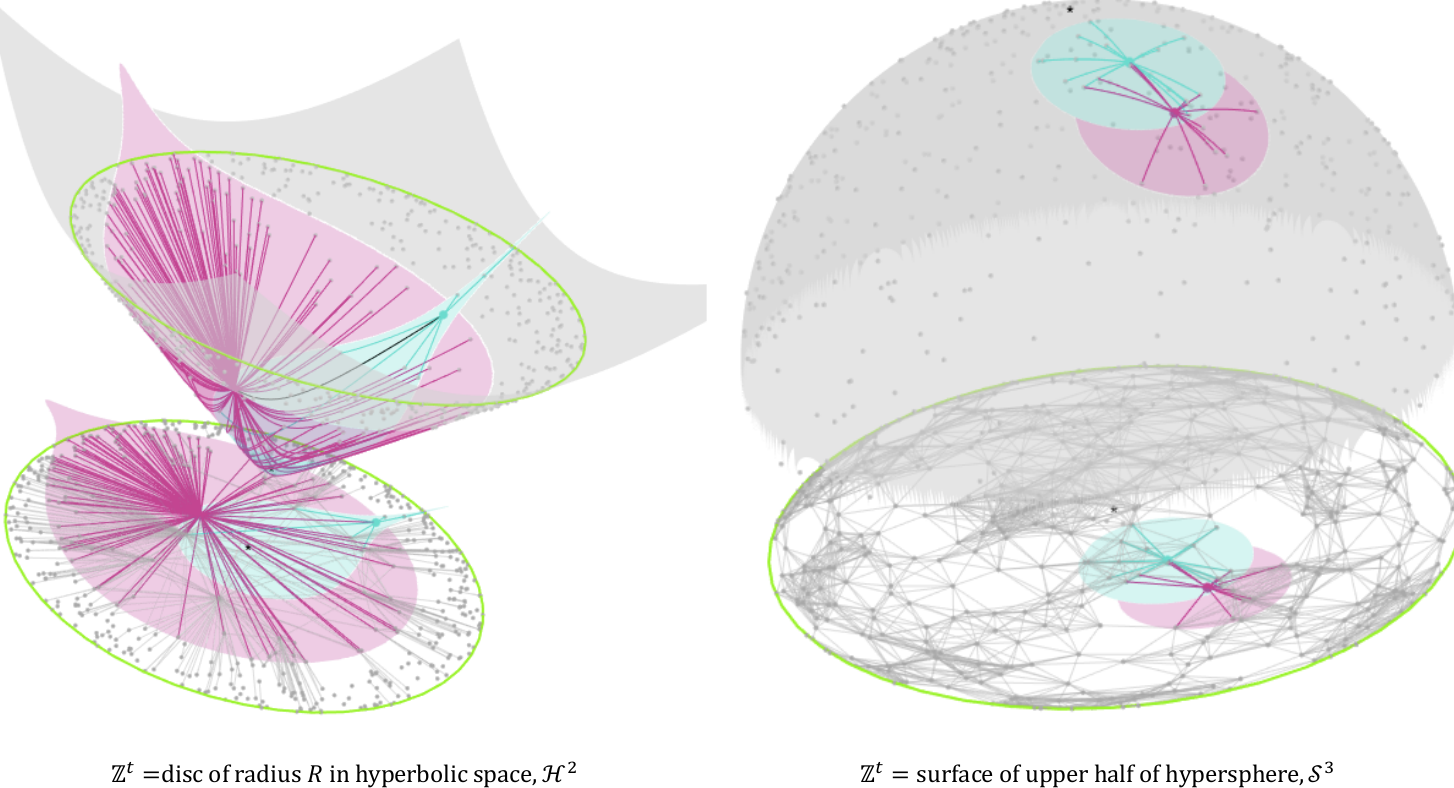}
\caption{Networks Simulated in Complex Geometries ($n=500$ nodes).  In each panel, the nodes are displayed both in their natural latent space (on the surface of a hyperbola -- in theory, this space extends up to infinity -- or the surface of a sphere, each shaded in light grey) as well as projected onto the $xy$-plane (the green circles below each latent space).  In both models, latent positions are constrained to fall within a radius of $R$ from the origin;  the green circles identify these limiting bounds.  Additionally, we highlight two nodes by drawing circles of radius $R$ about them (in light pink and light teal) and color their connections to other nodes in the network (identified by the dark pink and dark teal lines).  All other nodes are displayed as grey points with their connections identified by gray lines in the $xy$-plane.  For hyperbolic space, $R$ is chosen such that the average degree is eight \citep[][eq. 13]{krioukov_papadopoulos_kitsak_etal_2010} and in elliptic space, $R=1$.}
\label{figBowl}
\end{figure}

In this sense, hyperbolic space can be thought as having ``more space'' than Euclidean space and this fact has also been used to help visualize large networks \citep{lamping_rao_pirolli_1995,munzner_1997}.  For example, visualizing a network in the Poincar\'{e} disk (see Appendix \ref{HypModels.sec} and Figure \ref{figHypModels} for a more detailed description of the Poincar\'{e} disk) is like viewing the network through a fish-eye lens, with nodes at the center of the disk in focus and displayed with the most detail while nodes that are further away are pushed towards the boundary of the disk and are diminished in size\footnote{Recently, some video games have taken advantage of this feature of hyperbolic space as well, including the maze-like game developed by \citealp{madore_game} and the role playing game developed by \citealp{rogue_rogue_2011} and described here \citealp{celinska_kopczynski_2017};  both games have online demonstrations and can help provide some intuition for movement and distances within a hyperbolic space.}.  In fact, because hyperbolic space is ``bigger'' than Euclidean space, it cannot be exactly embedded within it.  Thus, unlike elliptic space where we can, for example, easily picture the surface of a hypersphere in $\mathbb{R}^3$, there exist many models for visualizing hyperbolic space, each of which highlights different features of hyperbolic geometry (see Appendix \ref{HypModels.sec} for more details).  One popular frame of reference for 2-dimensional hyperbolic space is the hyperboloid model where points on the hyperbolic plane are modeled as points on the surface of the upper sheet of a hyperboloid in $\mathbb{R}^3$.  This model can help us to visualize why hyperbolic space may be useful for understanding network data:  Consider any two individuals in a social network and draw the shortest path between them.  This path will most likely pass through some subset of individuals who are more centrally located in the network (e.g., inviduals that have many connections, particularly to other highly connected individuals).  Similarly, consider any two points on a hyperboloid;  the shortest path between them must pass through the ``more central'' points located in the bottom or bowl of the hyperboloid, those points nearest the origin (we develop this idea further in Figure \ref{figBowl}).

\citet{krioukov_papadopoulos_kitsak_etal_2010} also provide a higher-level argument based on the observed hierarchical structure of many real world networks \citep{clauset_moore_newman_2008}. More specifically, note that networks, by definition, connect heterogeneous nodes. Consider the familiar example of a social network, where the nodes represent individuals who may have different genders, ages, family backgrounds, etc. This heterogeneity implies that some sort of classification or grouping exists. We can imagine grouping the nodes of a social network first by gender, then by age group within gender, etc., creating a hidden hierarchical structure. Finally, the relationships among these groups could be represented as a tree-like structure. Thus, networks can be thought of as (even approximate) trees, which we have already described as living in hyperbolic space.  And in fact, many networks exhibit tree-like characteristics \citep[see, for example,][]{abu-ata_dragan_2016}.  Note that this line of thought mirrors the justification for the use of ultrametric space provided by \citet{schweinberger_snijders_2003}.  In fact, ultrametric spaces are very closely related to hyperbolic space \citep[see, for example][]{ibragimov_2014}.  However, in generic hyperbolic space, the natural distance metric does not satisfy the strong triangle inequality.

%\subsubsection{A Toy Example}
%\label{toySim.sec}

\renewcommand{\arraystretch}{2.25}
 \begin{table}[t]
 \caption{Geometric Properties of Curved Spaces.  This table is adapted from \citet{krioukov_papadopoulos_kitsak_etal_2010}.}
 \centering
\begin{tabular}{lll|ccc}
\hline \hline
&&& \multicolumn{3}{c}{GEOMETRY} \\
%\cline{4-6}
&&& \textbf{Euclidean} & \textbf{Spherical} & \textbf{Hyperbolic} \\
\hline \hline
\parbox[t]{1.5mm}{\multirow{7}{*}{\rotatebox[origin=c]{90}{PROPERTY}}}  &&Curvature, $K$ & $0$ & $>0$ & $<0$ \\
\cline{2-6}
&& \parbox[c]{1.5in}{Number of parallel lines} & 1 & 0 & $\infty$ \\
\cline{2-6}
&& \multirow{2}{*}{Triangles look$\dots$} & normal & thick & thin \\
&&& \includegraphics[width=.33in,]{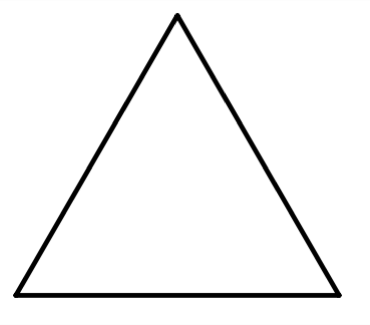} & \includegraphics[width=.33in]{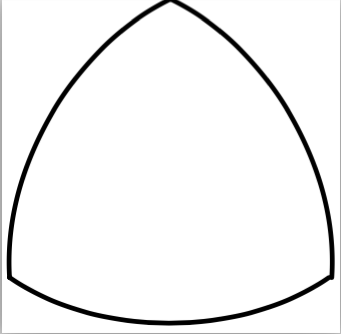} & \includegraphics[width=.33in]{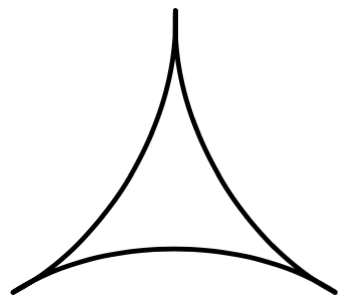} \\
\cline{2-6}
&& \parbox[c]{1.5in}{Sum of angles in a triangle} & $\pi$ & $> \pi$ & $< \pi$ \\
\cline{2-6}
&& \parbox[c]{1.5in}{Circumference of a circle} & \scriptsize{$2 \pi r$} & \scriptsize{$2 \pi \text{sin} \left( \sqrt{|K|} r \right)$} &  \scriptsize{$2 \pi \text{sinh} \left( \sqrt{|K|} r \right)$} \\
&& $\hspace{8mm} \dots$as $r$ increases, & \parbox[c]{.75in}{\centering linear growth} & \parbox[c]{.75in}{\centering sublinear growth} & \parbox[c]{.75in}{\centering exponential growth} \\
\hline \hline
%&&Area of a disk & $2\pi r^2 / 2$ & $2 \pi (1-\text{cos} \sqrt{|K|} r)$ & $2 \pi (\text{cosh} \sqrt{|K|} r - 1)$ \\
\end{tabular}
\label{tabCurve}
\end{table}
\renewcommand{\arraystretch}{1}

The key aspect of hyperbolic space which allows for a tree's exponential branching and which can accommodate complex network structure is its negative curvature. This directly influences the behavior of familiar geometric properties in hyperbolic space.  For example, triangles in hyperbolic space appear skinny or thin, with the sum of their angles being less than 180$^{\circ}$ (see Table \ref{tabCurve}). Similarly, distances in hyperbolic space behave differently than those in Euclidean space. For example, in Figure \ref{figBowl} we simulate 500 points in 2-dimensional hyperbolic space (using the hyperboloid model of $\mathbb{H}^2$ for visualization and simulating points uniformly\footnote{See equation \ref{eqnKri} and the following discussion for a more detailed description of the uniform distribution for discs in hyperbolic space.} within a disk of radius $R$), and connect all points within $\gamma=R$ units of each other.   This is equivalent to replacing the logit link function in model \ref{hypModel.eq} with a Heaviside step function, $p_{ij} = \Theta \left\{ \gamma-d(z_i,z_j) \right\}$, where the Heaviside step function is defined by
\[ \Theta(x) =  \left\{ \begin{array} {lr}
        1 & x \ge 0\\
        0 & \text{otherwise.}
        \end{array}\right.  \]
Of course, rather than deterministically connecting points within a certain distance, we could easily use the original model outlined in \ref{hypModel.eq}.  However, although the probabilistic link function allows for greater variation in simulated networks,  the Heaviside step function instead emphasizes the role of the distance metric in the latent space network model.  In the left panel of Figure \ref{figBowl}, we have also drawn circles of radius $\gamma=R$ about two particular simulated points.  First, it should be clear that distances behave differently in hyperbolic space, since these circles look very different than what we would expect in Euclidean space.  Second, note that most of the lines or connections are radially oriented. In fact, if we imagine this simple simulation as a toy latent space network model, we see that this tendency towards radially oriented connections might create networks that have heterogeneous degree distributions and are more centralized. Points or nodes near the base of the hyperbola are the more central nodes in the network, with lots of connections to nodes further from the origin, near the periphery. These peripheral nodes have fewer connections to the less central nodes along the periphery (fewer connections to other nodes that are also far from the origin) and are most likely connected to the central nodes near the origin. This type of centralization is often observed in real world networks and, for the latent space model, is simply an artifact of the negative curvature of the latent hyperbolic space.

Similarly, we can mimic this construction in positively-curved space, embedding nodes on the surface of a $3$-dimensional hypersphere as suggested by \citet{hoff_raftery_handcock_2002} and \citet{mccormick_zheng_2015}.  In the right panel of Figure \ref{figBowl} we simulate 500 points on the surface of the upper half of a 2-dimensional hypersphere (naturally visualized in 3-dimensional Euclidean space), again simulating points uniformly across this space.  Again, we use the toy version of the latent distance model which incorporates the Heaviside step function, so that all points within $\gamma=R$ units of each other are connected.  Note that the hyperbolic space's striking pattern of radially oriented ties does not appear in positively-curved space, indicating that we will not see much centralization or heterogeneity of the degree distribution.  Instead, the network simulated on the surface of the hypersphere appears to have more community structure, with sets of nodes that are highly interconnected but have fewer connections to other nodes outside of their set.  To our knowledge, there has been no formal development of theoretical or practical reasons as to why positively-curved space might be particularly amenable to latent space models for network data.  However, in \citet{mccormick_zheng_2015}'s application of this model to aggregated relational data, the authors mention a few appealing characteristics of working on the surface of a hypersphere.  First, calculations are simplified by the fact that the space is finite so that there is no need to restrict latent positions to some subset of the space (i.e., a (finite) disk of radius $R$ in (infinite) hyperbolic space).  Second, because the space is finite, this provides a natural non-zero lower bound for the probability of a tie in the network.  This assumption is more amenable to many practical settings, especially for aggregated relational data since the probability that an individual is tied to at least one member of a relatively large group of individuals is likely nonzero in practice.

\begin{figure}[t]
%trim = left bottom right top
\vspace{-.1in}
\begin{tabularx}{\textwidth}{ XXX }
\center{$\hspace{5mm} \alpha=0.5$} & \center{$\hspace{5mm} \alpha=1$} & \center{$\hspace{5mm} \alpha=2$}
\end{tabularx}
\vspace{0.05in}

\includegraphics[width=1\textwidth,trim=0in .1in 0in 0in, clip]{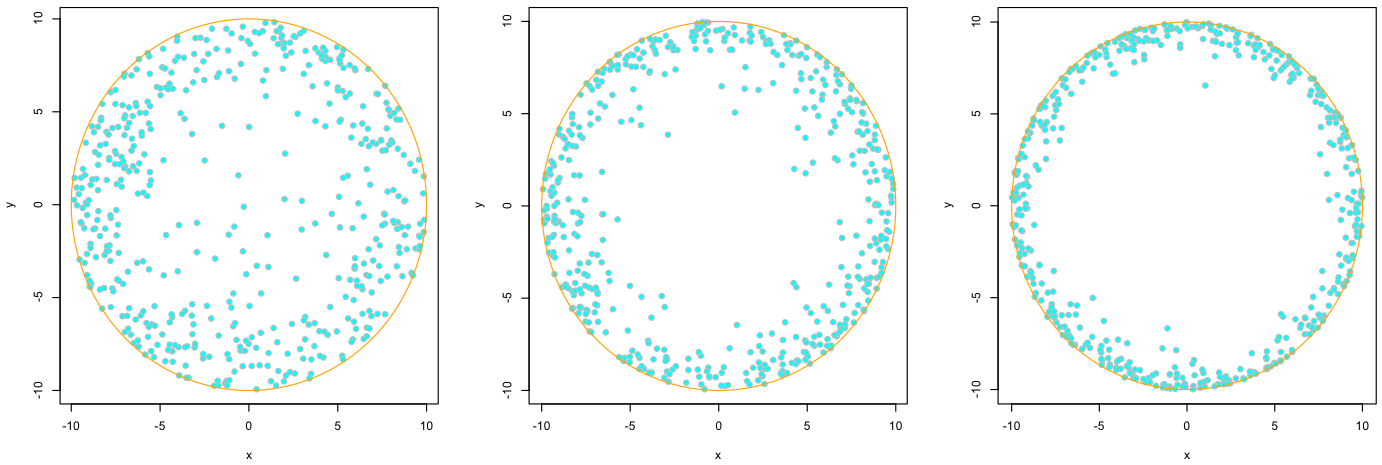}

\caption{Quasi-uniform Distribution of Latent Positions in Hyperbolic Space.  Although the distribution for the latent positions embeds nodes in hyperbolic space, here we plot only the $x$- and $y$-coordinates for these simulated positions, so that the plots above are in $\mathbb{R}^2$, for simplicity.  When $\alpha=1$, the points are exactly uniformly distributed and we can see that as $\alpha$ increases, points become more clustered around the boundary.}

\label{figRadii}
\end{figure}

\citet{krioukov_papadopoulos_kitsak_etal_2010} derive a hyperbolic latent space model which directly connects model parameters to important features of network structure, but departs from our specification of a generic latent space model in \ref{eqLSMgen}.  The authors outline the following model in $\mathbb{H}^2$:
\begin{align} \label{eqnKri}
Y_{ij} & \overset{ind}{\sim} \text{Bernoulli}(p_{ij}) \nonumber \\
\text{logit}(p_{ij}) &= \frac{1}{2T} \left\{ R(n,\bar{k},\gamma,T) - d(\bm{z}_i, \bm{z}_j) \right\}\\
r_i & \overset{iid}{\sim} p(r|R,\alpha(\gamma,T)),\>\> \phi_i  \overset{iid}{\sim} \text{Uniform}(0,2 \pi), \nonumber
\end{align}
where the latent positions are specified in polar coordinates, $\bm{z}_i =(r_i,\phi_i)$ and distance between the $i$th and $j$th latent position is given by \ref{hypDist.eq}.  The latent positions are assumed to be quasi-uniformly distributed within a disk of radius $R$ centered at the origin,
\[ p(r|R,\alpha) = \alpha \frac{ \text{sinh}(\alpha \> r) }{ \text{cosh}(\alpha R)-1 }, \hspace{7.5mm} \alpha = \left\{ \begin{array} {lr}
        \frac{1}{2}(\gamma-1), & T \le 1\\
        \frac{1}{2T}(\gamma-1), & T > 1,
        \end{array}\right. \]
which is roughly equivalent to an Exponential($\alpha$) distribution.  Note that $p(r|R,\alpha)$ distributes points exactly uniformly within a disk of radius $R$ when $\alpha=1$ (See Figure \ref{figRadii} for examples of other values of $\alpha$).  In the model, the radius of this disk is a function of expected network characteristics, $R \equiv R(n,\bar{k},\gamma,T)$ where $\bar{k}$ is the expected average degree, $\gamma$ is the exponent of the power-law degree distribution, and $T$ controls the amount of clustering, and satisfies the following equation:
\[ \bar{k} = \frac{N}{\pi} \int_0^R p(r_1|R,\alpha) \int_0^R p(r_2|R,\alpha) \int_0^{\pi} \text{logit}(p_{ij}) \> d \phi_1 dr_1 dr_2.\]
In practice, numerical integration is used to calculate $R$ which complicates potential inference for this particular specification of the model.   Note that this formulation is similar to the generic model outlined in \ref{eqLSMgen}, with the exception of the baseline tendency for tie formation, $\alpha$.  Here, model \ref{eqLSMgen}'s $\alpha$ is given a very specific form and in fact depends on an important feature of the latent space - its radius or bound, $R$ - which itself is a function of parameters which correspond to important features of network data.  In this sense, model \ref{eqLSMgen}'s parameter for a baseline tendency for tie formation has been replaced by a term that represents the complicated interplay between features of the latent space and expected network properties.

%%%%%%%%%%%%%%%%%%%%%%%%%%%%%%%%%%
%%
%%    (  4  )   IMPLICATIONS OF LATENT SPACE GEOMETRY ON
%%			NETWORK SUMMARY STATISTICS
%%
%%%%%%%%%%%%%%%%%%%%%%%%%%%%%%%%%%
\section{Implications of latent space geometry on network summary statistics} \label{secStats}

Of course, \citet{krioukov_papadopoulos_kitsak_etal_2010}'s line of reasoning - that the negative curvature of hyperbolic geometry accommodates complex network structure - leads to a rather natural question:  what role does the curvature of the latent space play in latent space models for network data?  By considering networks simulated from latent distance models in spaces of varying curvature, we will explore the function of geometric curvature in the latent space model.  Further, we show that, for the purposes of building a hierarchical network model, changing the geometry of the latent space can parsimoniously grow the complexity of the latent space network model while maintaining the intuitive appeal of using distance to model dependence.

%For consistency, we will hold the distribution of latent positions constant across all spaces.  Here, we have chosen to assume that the latent positions are uniformly distributed within a disk of radius $R$, where $R$ is specified such that all three latent spaces, $\mathcal{Z}^t$, have the same diameter (maximum distance between pairs of points).  This corresponds to the latent spaces plotted in Figure \ref{figSpaceCompare}, where each space has the same intrinsic radius ($R=\pi/2$) and the curvatures of the curved spaces are comparable (i.e., setting $r=1$ defines an elliptic space with curvature +1 and similarly, the hyperbolic space has curvature -1).  For any geometric space, the intrinsic radius determines this curvature and other geometric properties of the space, independent of any coordinate system used to represent the spaces.  Note that this radius lies within each latent geometry (i.e., in Figure \ref{figSpaceCompare}, the intrinsic radius is a line along the surface of the sphere in elliptic space and along the surface of the hyperboloid in hyperbolic space) and controls how much of each geometric space is being used in the model;  latent space positions are sampled only within the disks of radius $R$.  

\begin{figure}[t]
%trim = left bottom right top
%\includegraphics[width=1\textwidth,trim=0 2.5cm 0 0, clip]{CompareSpaces.eps}
\includegraphics[width=1\textwidth]{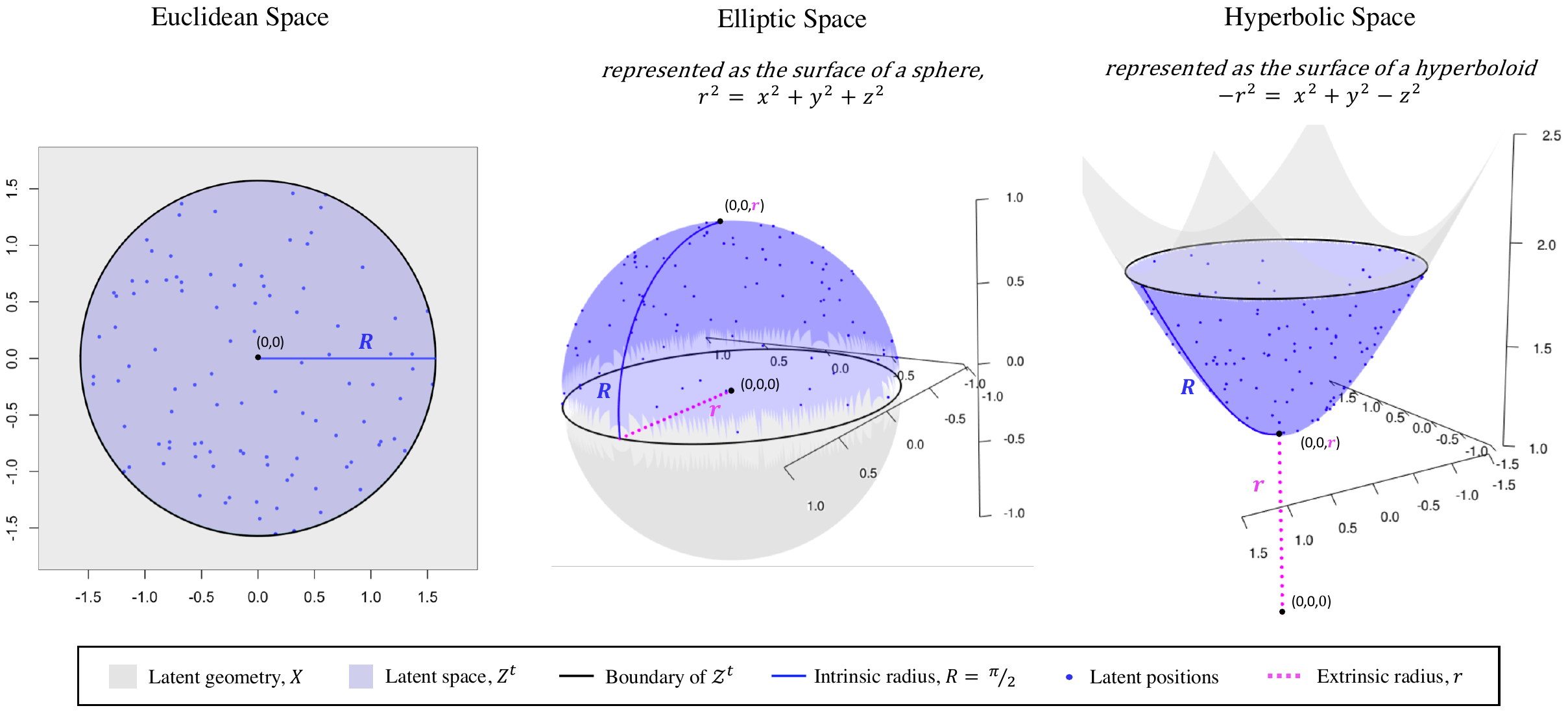}
\caption{Comparable Latent Spaces.  Latent positions are sampled uniformly from a subset (here, $Z^t$ is defined as a disc of radius $R=\pi / 2$) of each latent geometry, with elliptic geometry represented as the surface of a sphere and hyperbolic geometry represented by the surface of a hyperboloid plotted in $\mathbb{R}^3$.  The intrinsic radius, $R$, of the disc controls how much of each latent space, $X$, is being used in the model as $Z^t$.  This should not be confused with the extrinsic radius, $r$, which determines the curvature of elliptic and hyperbolic space (here, $r=1$ and $r=-1$, respectively).}
\label{figSpaceCompare}
\end{figure}

First, note that the curvature of a space is determined by the extrinsic radius, $r$;  here, we are examining a hyperbolic space with curvature $r=-1$, a Euclidean space with curvature $r=0$, and an elliptic space with curvature $r=1$ (see Figure \ref{figSpaceCompare}).
For consistency, we will define $Z^t$ to be a disk with intrinsic radius, $R$, centered at the origin (or its analogue\footnote{To be concrete, in Euclidean space the disk is centered about (0,0).  Using the models of elliptic and hyperbolic space in $\mathbb{R}^3$ (see Figure \ref{figSpaceCompare}), the disks in elliptic and hyperbolic space are each centered about (0,0,1).}) in each of the three latent spaces and will assume that the latent positions are uniformly distributed within this disk.
For any geometric space, $X$, a disk with intrinsic radius $R$ is the set of all points in $X$ having distance less than or equal to $R$ from a specific fixed point.
As a result, all three latent spaces, $Z^t$, have the same diameter (maximum distance between pairs of points).
This corresponds to the latent spaces plotted in Figure \ref{figSpaceCompare}, each of which has the same intrinsic radius but different curvatures.
To better visualize these spaces, we have drawn Figure \ref{figSpaceCompare} with intrinsic radius $R=\pi/2$, but in the some of the simulations that follow we will use $R=\pi$, resulting in an elliptic latent space that is the entire sphere.
Note that $R$ controls how much of each latent space, $X$, is being used in the model as $Z^t$;  latent positions are sampled only within disks of intrinsic radius $R$.

\begin{figure}[t]
%trim = left bottom right top
%\includegraphics[width=1\textwidth,page=4,trim=0.65in 1.5in 0.65in 1.5in, clip]{SideBySides}
\includegraphics[width=1\textwidth]{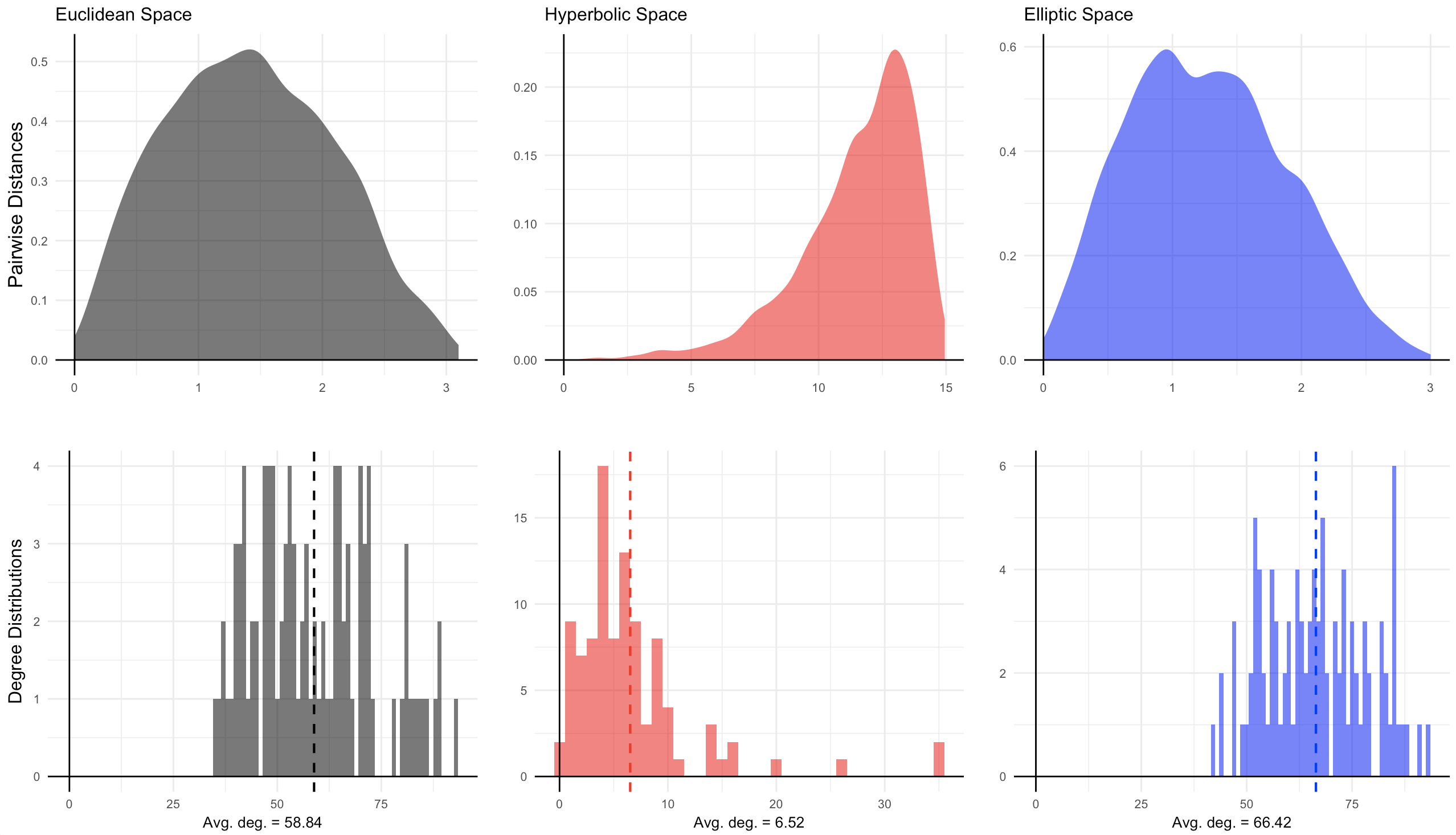}

\hspace{6mm}
\begin{minipage}[b]{0.27\linewidth}
\includegraphics[width=\textwidth]{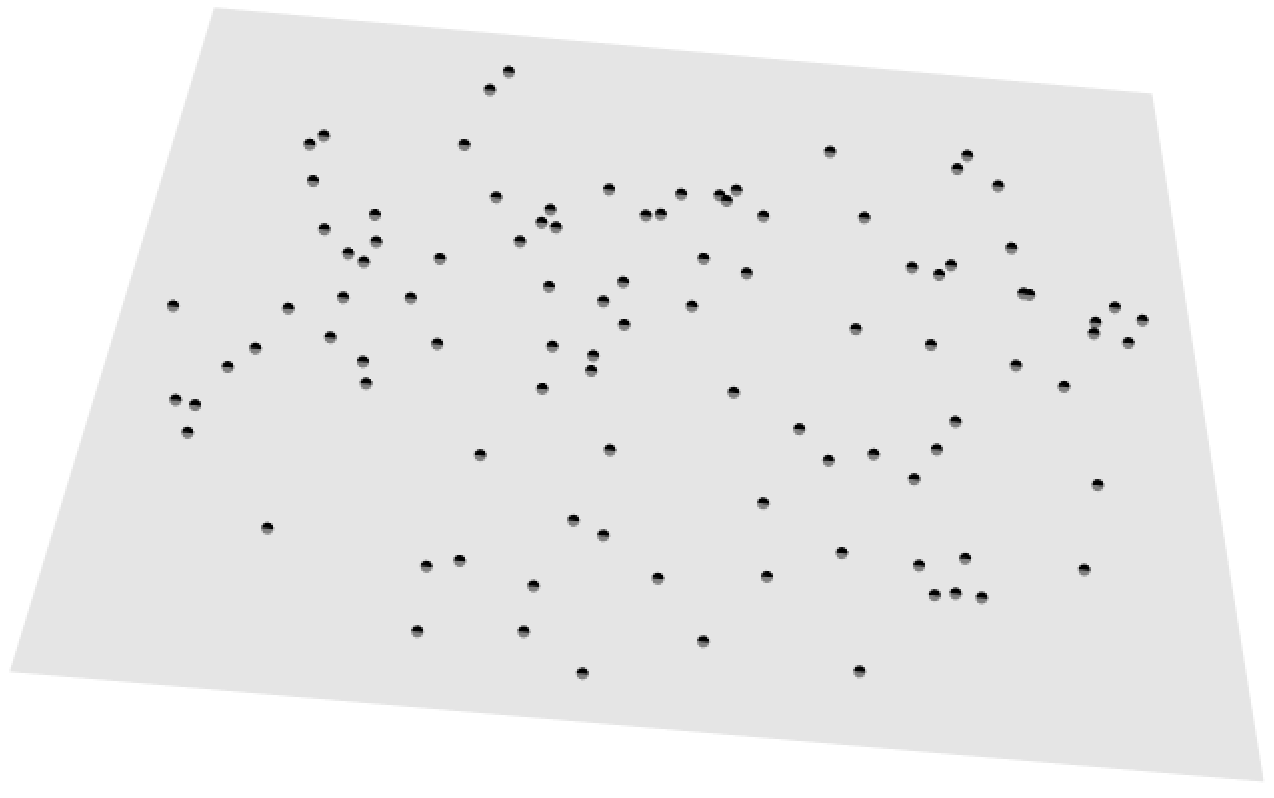}
\vspace{0.5mm}
\end{minipage}
\hspace{8mm}
\begin{minipage}[b]{0.27\linewidth}
\includegraphics[width=\textwidth]{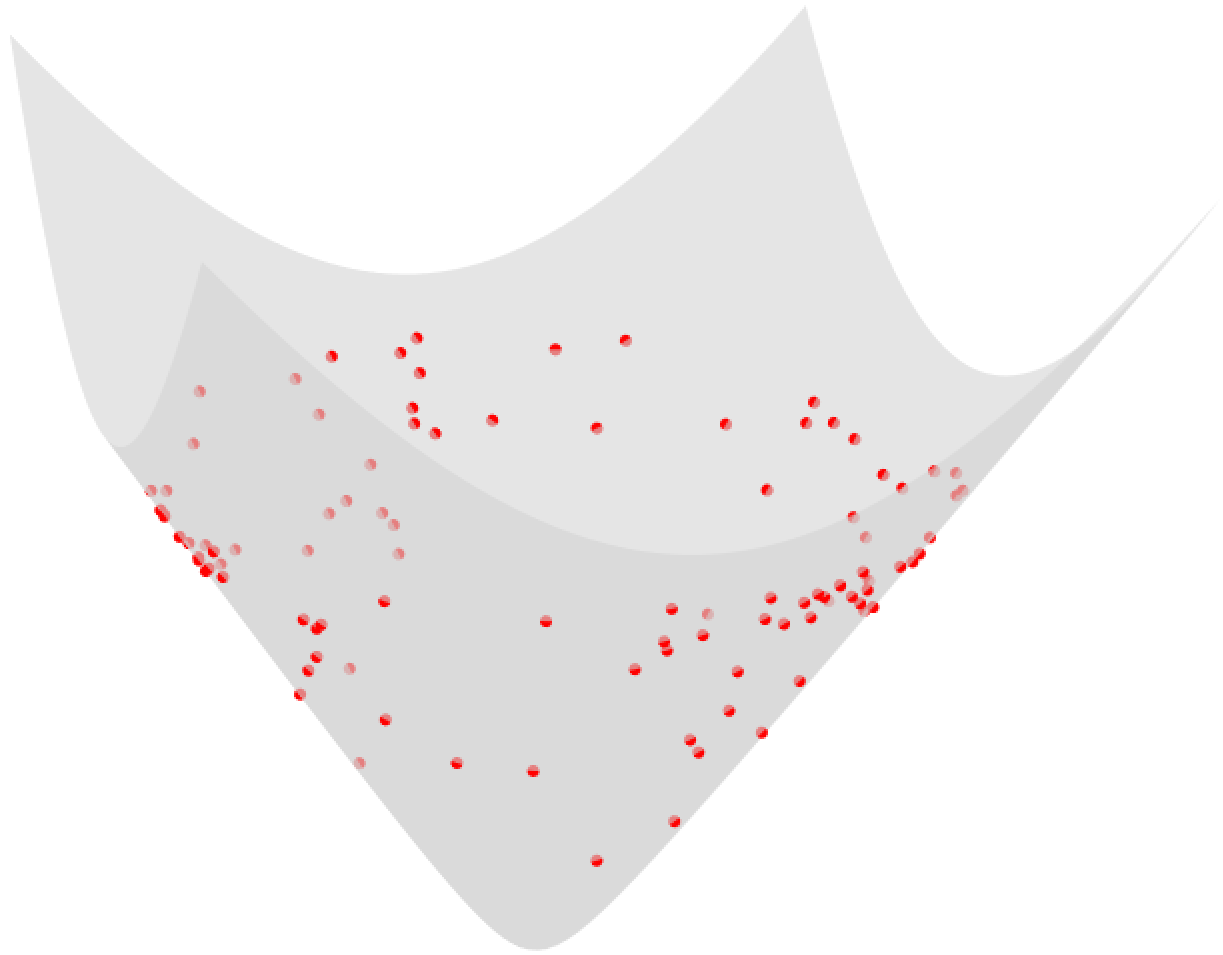}
\end{minipage}
\hspace{3mm}
\begin{minipage}[b]{0.27\linewidth}
\includegraphics[width=\textwidth]{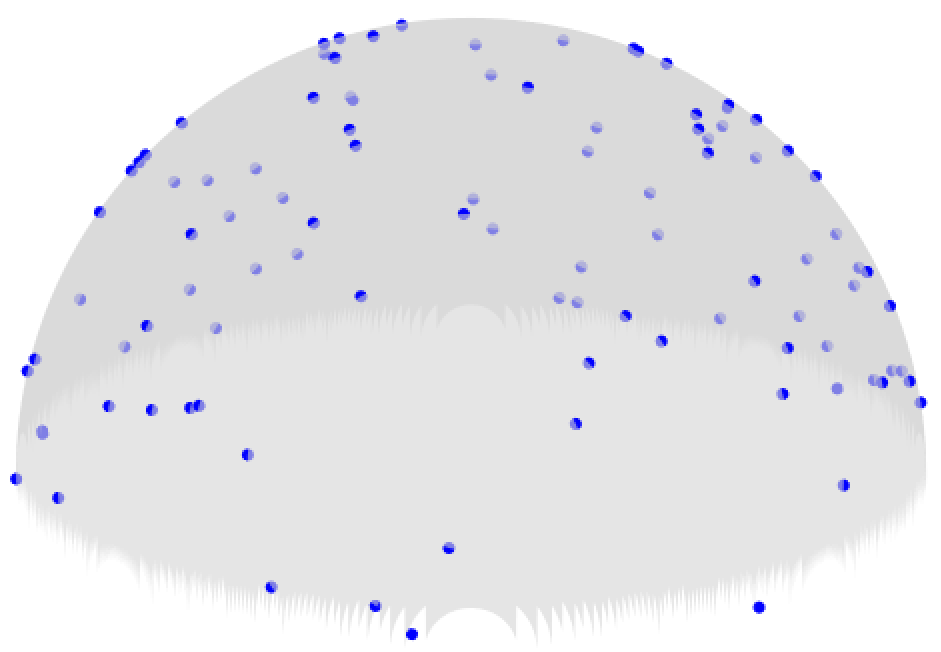}
\end{minipage}

\caption{Distribution of Distances in Latent Spaces with Differing Geometries.  Panels in the bottom row plot one set of $n=100$ points simulated in Euclidean, hyperbolic, and elliptic space, respectively, while panels in the top row plot histograms of the pairwise distances for these sets of points.  The center row plots the corresponding degree distributions across the three latent geometries, where nodes in each network are connected according to model \ref{simMod.eq} with $\gamma=R$.}
\label{figDistHist}
\end{figure}

\begin{figure}[t]
  \centering
  \includegraphics[width=\textwidth,trim = 0 .5cm 0 0cm, clip]{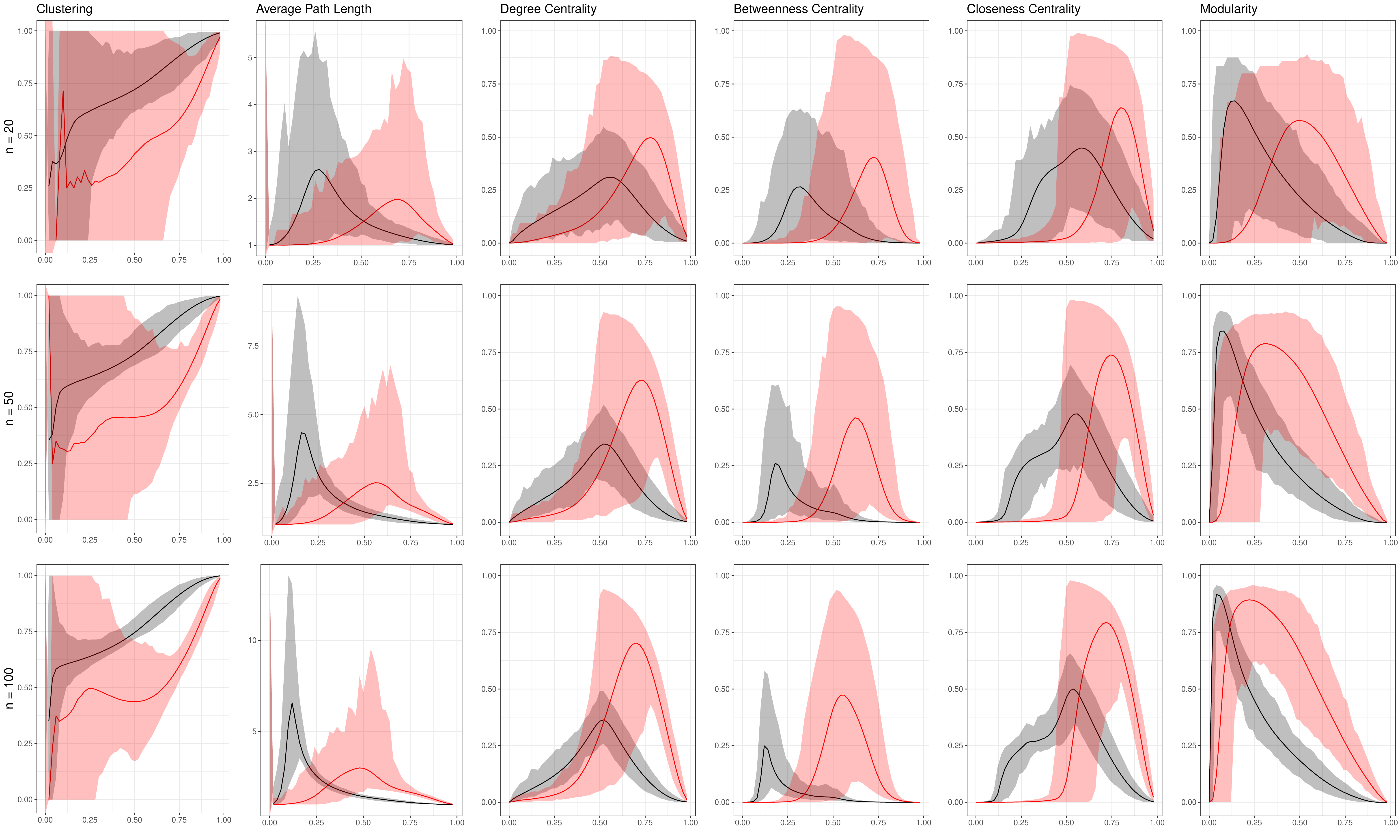}
  \includegraphics[width=.35\textwidth,trim=0 2cm 0 0,clip]{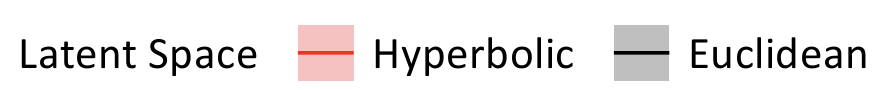}
  %\vspace{-.25in}
  \caption{Network Statistics in Hyperbolic Space.  The $x$-axis is $\gamma$, the Heaviside step function cut point.  Distances across the geometries are made comparable by re-scaling each simulation to the $[0,1]$ interval.  Each row of plots corresponds to simulated networks of a different size, $n=20, 50, 100$, and each column of plots corresponds to a different network summary measure.}
  \label{figDistBands}
\end{figure}

\begin{figure}[t]
  \centering
  \includegraphics[width=\textwidth]{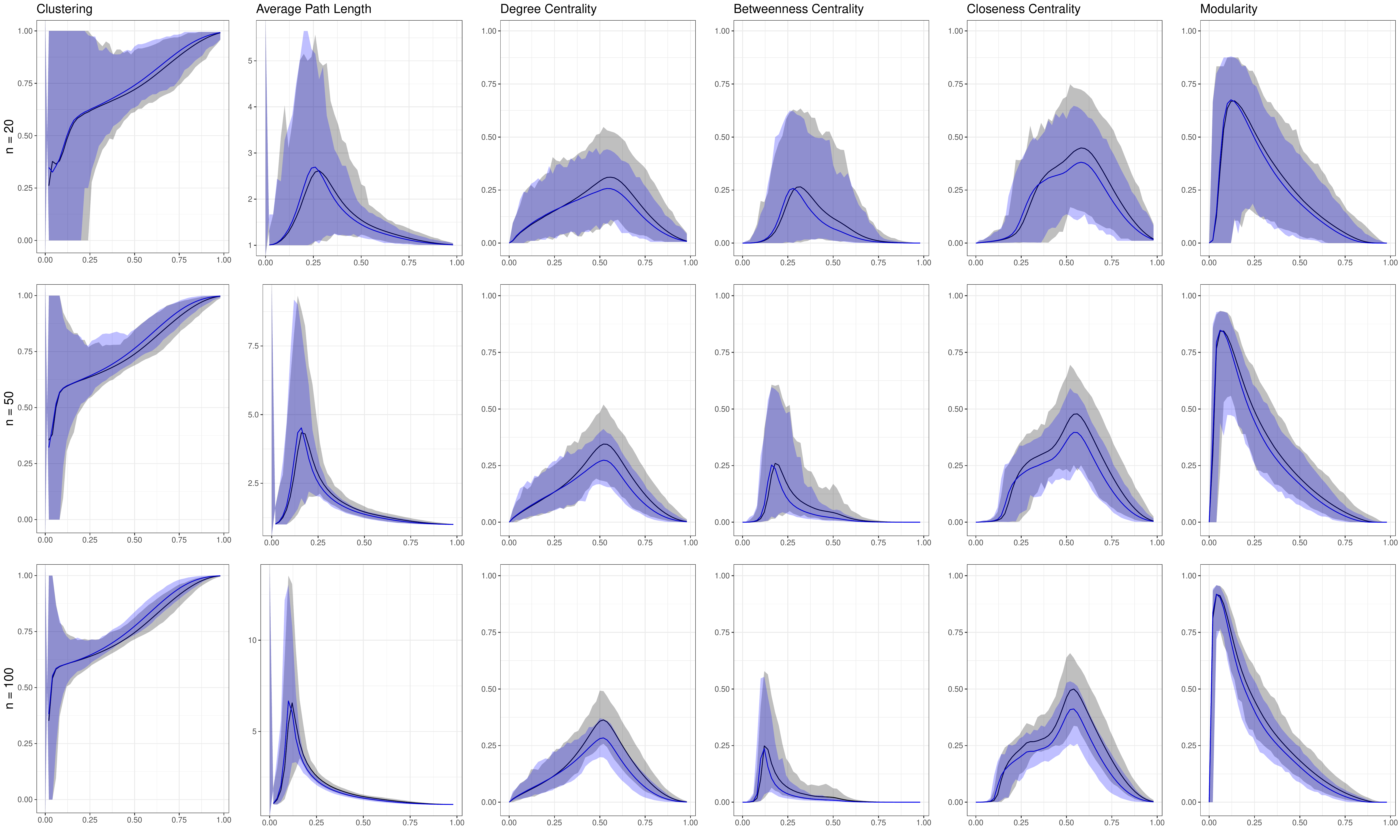}
  \includegraphics[width=.35\textwidth,trim=0 0.25cm 0 2cm,clip]{distBandsLegends}
  %\vspace{-.25in}
  \caption{Network Statistics in Elliptic Space.  The $x$-axis is $\gamma$, the Heaviside step function cut point.  Distances across the geometries are made comparable by re-scaling each simulation to the $[0,1]$ interval.  Each row of plots corresponds to simulated networks of a different size, $n=20, 50, 100$, and each column of plots corresponds to a different network summary measure.}
  \label{figDistBands2}
\end{figure}

Specifying latent spaces that are comparable across the three geometries is essential to ensuring fair comparisons of any resulting network features.  However, doing so has required that we deviate from some of the standard assumptions of popular CLS models;  for example, in the \citet{hoff_raftery_handcock_2002} model, latent positions are drawn from a bivariate standard normal distribution.  The relationship between the \textit{distribution} of the latent positions in a given latent space (as opposed to the \textit{geometry} of the latent space) and emergent network features is not the focus of this paper and remains an interesting topic for further research.  We will rely on the comparable latent spaces defined above and depicted in Figure \ref{figSpaceCompare} (with some modification to the value of $R$) in the simulations described in this section (and depicted in Figures \ref{figDistHist}, \ref{figDistBands}, and \ref{figDistBands2}) as well as in Section \ref{secLaPlac} (and Figure \ref{figRealWorld} included there as well as Figure \ref{figSim} in the Appendix).  %To begin, to better illustrate the differences between these geometric spaces, in the simulations described in this section we will increase the intrinsic radius to $R=10$ in hyperbolic space (leaving $R=\pi/2$ in Euclidean and elliptic space), since many interesting features of this space are more striking further away from the origin.

Although we might consider directly examining simulated networks under each of these models, binary data can be particularly difficult to study and binary relational data (i.e., networks) are no exception.  For example, in the latent space model, even if $p_{ij}$ smoothly varies across potential ties in the network, this smoothness can not be manifested in the observed network, since observed ties are by definition either present or absent.  Thus, in our simulation study, we use a Heaviside step link function as was used in Figure \ref{figBowl}, again to emphasize the roles of the different distance metrics in each latent geometry.  Here, we consider smoothly varying $\gamma$, the Heaviside step function cut point, and examining differences in network structure across three latent geometries:  Euclidean space, hyperbolic space, and elliptic space.  The model used in this simulation exercise, is thus as follows:
\begin{equation} \begin{aligned}
Y_{ij} &= \Theta \left\{ \gamma - d(z_i,z_j) \right\} & i &< j; \>\> i,j=1,...,n \\
z_i & \in \mathcal{Z}^t, \>\> z_i \overset{iid}{\sim} \text{Uniform}(R) & i & = 1,...,n, \label{simMod.eq}
\end{aligned} \end{equation}
which is equivalent to replacing the logit link function in model \ref{hypModel.eq} with the Heaviside step function.  In this case, the tie variables, $Y_{ij}$, can still technically be written as Bernoulli variables but are actually deterministically, rather than probabilistically, specified given the set of latent positions.  Of course, while these choices greatly simplify the model originally outlined by \ref{hypModel.eq}, they allow us to more directly consider the role of geometry in the latent space model for network data.

First, it is worth noting that the typical distribution of distances varies greatly across these geometries.  For example, consider distances from a set of points simulated under each model\footnote{Variation across simulations does not appear to impact the conclusions drawn here.} in Figure \ref{figDistHist}.  %In Euclidean space, the distribution of distances is skewed right, with most points close together and only a few points very far apart.
The distribution of distances for latent positions on the hyperboloid looks very different than for those in the other spaces.  In hyperbolic space, there are few points that are very close together and the distribution of distances is left skewed.  This distribution likely better represents our intuition about the distances between latent positions for nodes in real world social networks - any randomly selected pair of individuals is not likely to have a tie (i.e., any pair of positions is likely far apart).  In both Euclidean\footnote{Although not depicted here, using a standard normal distribution (or even altering the standard deviation so that the maximum expected distance between latent positions mimics that of elliptic space) in Euclidean space \citep[as in][]{hoff_raftery_handcock_2002} results in a distribution of pairwise distances that is right-skewed and a degree distribution that implies that most individuals are very social, but there may be a few actors with very few ties.} and elliptic space, the distribution of distances is much more symmetric.  Recall that we have chosen the same distribution for the latent positions (uniform within a disk of radius $R$) across all three geometries, so that observed differences in the distributions of the distances in Figure \ref{figDistHist} is due to the differing curvature of the spaces.  However, to better illustrate the differences between these geometric spaces, in the simulations described in this section we have increased the intrinsic radius to $R=7.5$ in hyperbolic space (leaving $R=\pi/2$ in Euclidean and elliptic space\footnote{Note that in elliptic space,  $R$ is bounded above by $\pi$, so increasing $R$ and keeping $R$ equal across all three geometries is not possible}), since many interesting features of this space are more striking further away from the origin.

Naturally, one can imagine how these differences in the distribution of distances might affect a network's degree distribution (see the center row in Figure \ref{figDistHist}).  In the social network setting, %a Euclidean latent space implies that most individuals are very social, but there may be a few loners (a few actors with very few ties);
a hyperbolic space implies that most individuals are relatively asocial with only a few very popular actors; while a Euclidean or elliptic space implies that there are some loners (some actors with very few ties) and some popular actors, but most actors display roughly similar levels of sociality.  Of course, our description of the intuitive degree distribution in a hyperbolic latent space is a natural way of describing degree heterogeneity, a long observed trait of many real world networks.  

The results of our simulation study where we vary the cut point in the Heaviside step function, $\gamma$, and examine changes in network features is displayed in Figures \ref{figDistBands} and \ref{figDistBands2}.  Even after specifying comparable latent spaces\footnote{As for Figure \ref{figDistHist}, we again use an increased $R=7.5$ in hyperbolic space in order to fully capture the features of this space.}, we still observe some variation in the range of pairwise distances across the latent geometries (consider the $x$-axes of the density plots for Euclidean and elliptic space, where $R=\pi/2$ in each case, in Figure \ref{figDistHist}).  In our simulation exercise, we accommodate for this difference in scaling by dividing the distances by the maximum simulated distance in each simulation, so that all distances are contained within $[0,1]$.  As a result of this scaling, note that as $\gamma$ increases, the density of each network will increase according to a smooth linear trend which is uniform across all three spaces.  This is a result of the fact that the latent positions in each space are uniformly distributed and ties in the network depend directly on the natural distance metric in each space.  In this sense, the scaling adjusts for the slight variation in the range of simulated pairwise distances and also results in networks that are comparable in terms of network density, for any particular $\gamma$.  We consider networks of various size ($n=20, 50, 100$) embedded in Euclidean, hyperbolic, and elliptic latent geometries.  For each geometry and network size considered, we simulate 5,000 networks according to model \ref{simMod.eq}, and slowly increase $\gamma$ from 0 to 1 by increments of 0.20.  We plot the average of common network summary measures\footnote{All network statistics were computed using the \texttt{igraph} \citep{igraph} and \texttt{network} \citep{network} packages in \texttt{R} \citep{citeR}.} - clustering, average path length, degree centrality, betweenness centrality, closeness centrality, and modularity\footnote{To provide a quick sense of any observed community structure, we use the \texttt{cluster\_edge\_betweenness} \citep{newman_girvan_2004} and \texttt{modularity} \citep{clauset_newman_moore_2004} function provided in the \texttt{igraph} \citep{igraph} package for \texttt{R} \citep{citeR}.  This modularity statistic is included simply to provide an exploratory look at the extent to which community structure may be observed across the latent geometries, particularly for elliptic space.  In practice, any reliable confirmatory conclusions would require more detailed consideration of the community detection model, including the expected number and size of the communities, and modularity measure for any particular set of network data.} - of the resulting simulated networks as a function of $\gamma$ in Figures \ref{figDistBands} and \ref{figDistBands2}.  Bands on these plots are pointwise minima and maxima across the simulations - the jaggedness of these bands is a consequence of finite size effects for the simulated networks.

In Figure \ref{figDistBands}, we compare networks in hyperbolic space to those in Euclidean space.  Note that networks in hyperbolic space can achieve much higher levels of degree centrality, betweenness, and closeness, and that this effect is more prominent for larger networks.  This coincides with our previous intuition for the behavior of the degree distribution of networks in hyperbolic space.  Note also, that this improved ability to model degree heterogeneity does not come at the expense of any dramatic losses in clustering or average path length. 
%In Figure \ref{figDistBands2}, while we do observe some increase in degree heterogeneity, this difference is not quite as exaggerated as in hyperbolic space (particularly for $n=100$) and overall we see that networks in elliptic space do not differ greatly from those in Euclidean space in terms of the other centralization measures.  Recall that in the simulated network on the surface of the hypersphere in Figure \ref{figBowl}, we observed some indications of a community structure, with distinct groups of nodes having more connections within the group and fewer connections across groups.  And in fact, we see some difference in the modularity statistics across the two spaces.
In Figure \ref{figDistBands2}, we see that networks in elliptic space do not differ greatly from those in Euclidean space in terms of these network statistics\footnote{Although not displayed here, there are some slight differences between Euclidean and elliptic space when latent positions are normally distributed within Euclidean space, as in the \citet{hoff_raftery_handcock_2002} model.  In this case, we observe some slight increases in the degree heterogeneity in elliptic space, but this difference is not quite as exaggerated as in hyperbolic space (particularly for $n=100$).  Recall that in the simulated network on the surface of the hypersphere in Figure \ref{figBowl}, we observed some indications of a community structure, with distinct groups of nodes having more connections within the group and fewer connections across groups.  And in fact, we see some difference in the modularity statistics across the two spaces.}.

%%%%%%%%%%%%%%%%%%%%%%%%%%%%%%%%%%
%%
%%    (  5  )   GRAPH LAPLACIANS
%%
%%%%%%%%%%%%%%%%%%%%%%%%%%%%%%%%%%
\section{Using graph Laplacians to identify an appropriate geometry} \label{secLaPlac}

As mentioned above, many common descriptive summary measures of network structure are sensitive to the size of the network.  Thus, while the behavior of these measures across latent spaces of different geometries can provide us with some indication of the role of geometry in these models, it is difficult to formally relate these statistics to properties of the latent space.  This issue is rooted in the fact that it is difficult to find a common language describing both networks and geometric spaces; such a common language is a prerequisite for comparing properties between the two sorts of structures.  However, as we will examine in this section, both the geometry of a space and the combinatorics of a graph or network can be described by linear operators whose eigenvalues are comparable.

For geometric spaces, the Laplace-Beltrami operator, $\nabla_M^2$, is an operator on smooth functions which map general Riemannian manifolds, $M$, to $\mathbb{R}$.  For a function, $f = f(x_1, x_2, \dots x_n)$, when $M = \mathbb{R}^n$, the Laplace-Beltrami operator can be expressed in terms of partial derivatives:
\begin{align}
\nabla^2_{\mathbb{R}^n} f = \sum_{i=1}^n \frac{ \partial^2 f}{ \partial x_i^2}, \label{LBop}
\end{align}
and hence when $M=\mathbb{R}$, the Laplace-Beltrami operator is simply the second derivative, $\nabla^2_{\mathbb{R}} f= f''$.  Even for general Riemannian manifolds, $M$, the Laplace-Beltrami operator, $\nabla_M^2$, is essentially defined by \ref{LBop} with respect to local coordinates $x_1,x_2,\ldots,x_n$.  The eigenvalues of the Laplace-Beltrami operator describe many (though not all) geometric properties of the space $M$.  The first positive eigenvalue of the Laplace-Beltrami operator is positively related to the curvatures of certain compact manifolds \citep{lichnerowicz_1958}.  More generally, a conjectural generalization of Weyl's Law \citep{weyl_1911} describes a relationship between the volumes of compact domains $D$ in any Riemmanian manifold $M$ and the eigenvalues of the Laplace-Beltrami operator $\nabla^2_M$, restricted to those functions with support $D$ \citep{ivrii_1980}.  Thus, in particular settings, we can describe a direct relationship between the eigenvalues of the Laplace-Beltrami operator and much of the geometry of a space.

The \textit{graph Laplacian} $\nabla^2_G$ of a finite graph $G$ is the operator on functions $V_G\rightarrow\mathbb{R}$, from nodes $V_G$ defined by the rule
\begin{equation*}
  (\nabla^2_G\phi)(v)=\deg(v)\phi(v)-\!\!\!\sum_{(v,w)\in E_G}\phi(w).
\end{equation*}
%When the vector space of functions $V_G\rightarrow\mathbb{R}$ is identified with $\mathbb{R}^{\#V_G}$, 
After ordering vertices from $1$ to $n$, $\nabla^2_G$ can be expressed by
\begin{equation*}
  \nabla^2_G=D_G-A_G,
\end{equation*}
where $A_G$ is the adjacency matrix of $G$ and $D_G$ is the diagonal matrix whose entries are the degrees of vertices of $G$.
%Variants of the graph Laplacian, such as the symmetric normalized graph Laplacian $I-D^{-1/2}_GA_GD^{-1/2}_G$, are not covered in this paper.
We note that that there are other versions of the graph Laplacian, including the symmetric normalized graph Laplacian. $I-D^{-1/2}_GA_GD^{-1/2}_G$, with which the reader may be familiar.

 \begin{figure}[t]
%trim = left bottom right top
\includegraphics[width=1\textwidth]{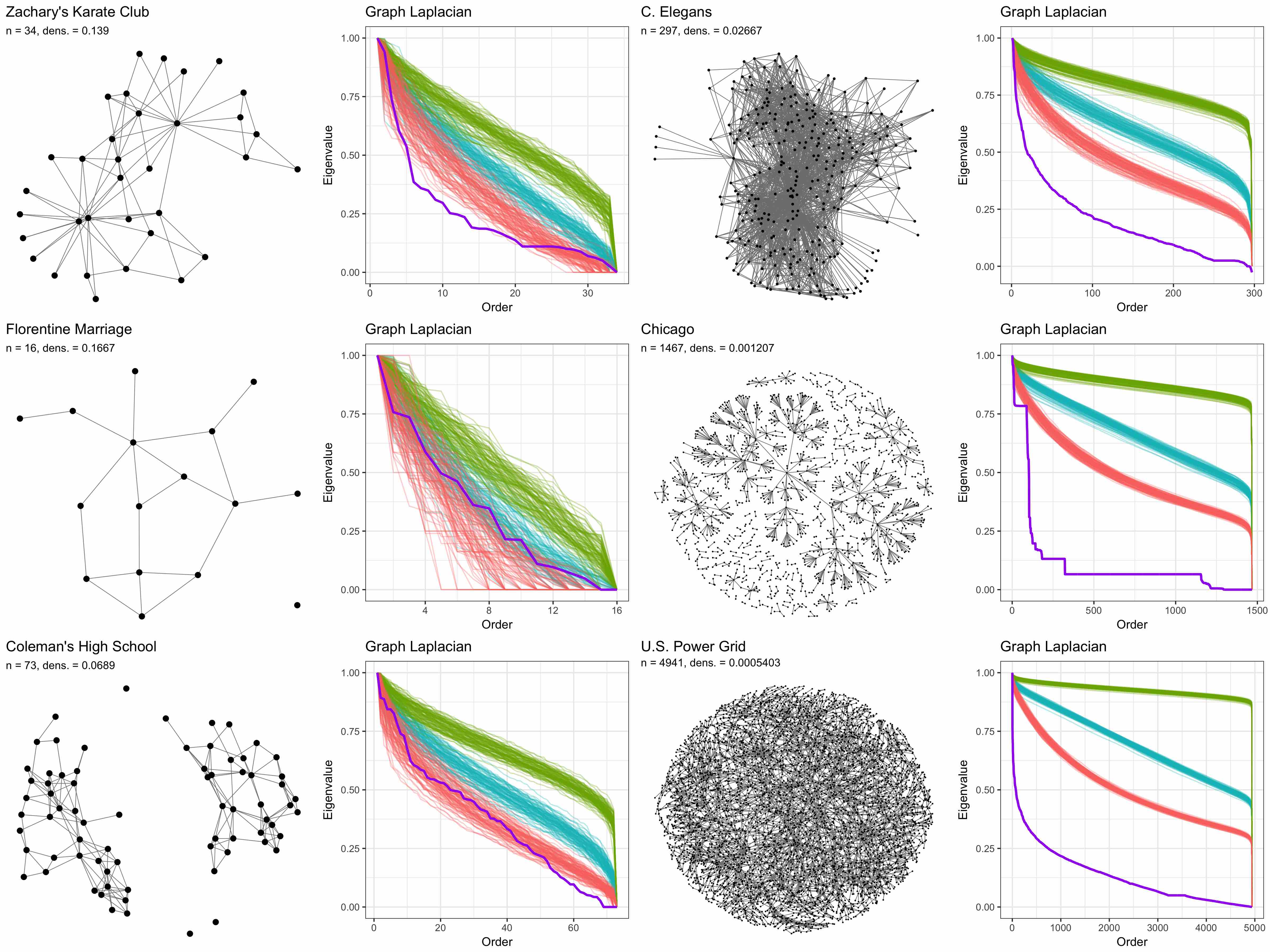}
\includegraphics[width=1\textwidth,trim={0 .5in 0 31in},clip]{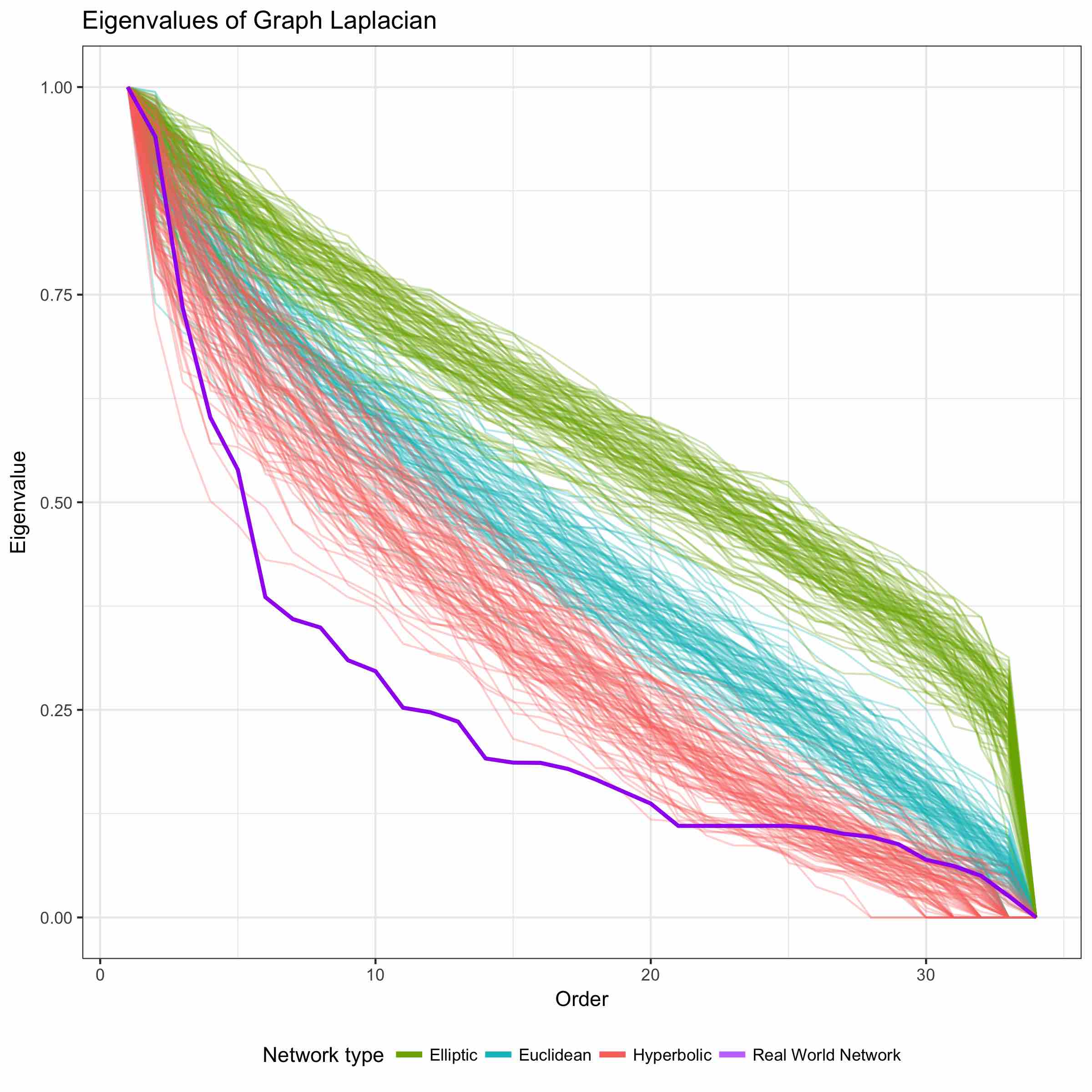}
\caption[Caption for LOF]{A small collection of real world networks.  In these plots, we compare a set of real world networks to networks simulated from our general latent distance model (Equation \ref{eqLDMgen}) across different geometries. Each row includes a network plot (using the default settings of the \texttt{ggnet} function in the \texttt{GGally} package for \texttt{R}) and a plot of the eigenvalue curves, where the color corresponds to the geometry of the latent space.  In the eigenvalue plots, each network is represented by a curve which plots the network's eigenvalues against their order.  Networks on the left side of the plot are examples of social networks, that range in size from $n=16$ to $n=73$, while networks on the right side of the plot are non-social networks, including a brain network, road networks and a power grid network.}
\label{figRealWorld}
\end{figure}

The graph Laplacian is a discrete analogue of the Laplace-Beltrami operator in the following sense.
Just as the Laplace-Belatrami generalizes the second derivative, the graph Laplacian generalizes a \textit{difference of differences}. 
For example, let $I$ be the discrete, graph-theoretic approximation
\begin{equation*}
  \xymatrix{
    v_1\ar@{-}[r] & v_2\ar@{-}[r] & \cdots\ar@{-}[r] & v_n,
  }
\end{equation*}
of the open interval $(0,1)$, where horizontal lines represent connections between the vertices of $I$, $v_i$ for $i = 1, \dots n$.
%\textcolor{red}{[Dena:  Thinking about the comment from Reviewer \#2, what about changing this to something like, ``of the open interval $(0,1)$, where $v_i$ are vertices on $(0,1)$ and horizontal lines represent connections between these vertices.''] [I'm worried that's more confusing than leaving the analogue with the graph above with (0,1) unsaid; after all, what would ``vertices on (0,1)'' technically mean?-Dena]}
Just as $\nabla^2_{(0,1)}f=f''$ , $\nabla^2_I\phi$ is defined on the inner ($1<i<n$) vertices $v_i$ by
$$(\nabla^2_I\phi)(v_i)=(\phi(v_{i})-\phi(v_{i-1}))-(\phi(v_{i+1})-\phi(v_{i})).$$

The eigenvalues of $\nabla^2_G$ describe many (though not all) combinatorial features of a graph $G$.
For example, the algebraic multiplicity of the eigenvalue $0$ reveals how many connected components $G$ has.
The second smallest eigenvalue of $\nabla^2_G$ quantifies how interconnected the nodes in $G$ are.

At least intuitively, the Laplace-Belatrami operator $\nabla^2_M$ is some sort of limiting approximation of graph Laplacians $\nabla^2_G$ of graphs $G$ sampled from a latent distance model whose latent space is the Riemmanian manifold $M$ .
Unfortunately, the closest sort of theoretical results of this nature \citep[c.f.][]{belkin_niyogi_2005, ting_huang_jordan_2011,hein_audibert_luxburg_2007} use a modification of $\nabla^2_G$ when the edges of $G$ are weighted by distances (e.g., Euclidean point-cloud data.) 

%\textcolor{red}{Need to update text in the following couple paragraphs and add a description of the experiments with intuitive network models (and the non-example of the lattice).}

We test this intuition with some simulations. Using the comparable latent spaces defined in Section \ref{secStats}, we consider models where nodes are uniformly distributed within a disk of radius $R = \pi$ in Euclidean, hyperbolic, and elliptic latent space.  %Here, we specify $R$ according to the approximate formula\footnote{$R \approx \text{log} \left( \frac{8n}{\pi \bar{k}} \right)$, where $n$ is the number of nodes and $\bar{k}$ is the average degree.} provided in \citep{krioukov_papadopoulos_kitsak_etal_2010} such that the simulated networks in hyperbolic space will mimic the average degree (or equivalently, density) of the examined real-world networks (maximum distance between pairs of points).
For each real world social network we examine, we generate 100 simulated networks from model \ref{eqLDMgen} (with $\alpha=0$), each with the same number of nodes as the real world social network, for each CLS model\footnote{In these simulations, we purposefully utilize a relatively simplistic CLS model in order to highlight differences due solely to the geometry of the latent spaces;  better model fit could most likely be obtained by using models that can incorporate more features specific to the real world network being considered, such as the \citet{hoff_raftery_handcock_2002} model or model \ref{eqnKri} \citep{krioukov_papadopoulos_kitsak_etal_2010}, for example.}.  %We consider networks that are popular in social network research, including Zachary's Karate Club network \citep{zachary_1977}, \textcolor{red}{list of other networks}. In the far left panel of Figure \ref{figLaPlacPic}, plots of the eigenvalues in decreasing order strongly suggest that Zachary's Karate Club network exhibits geometric features strongly reminiscent of hyperbolic space. \textcolor{red}{more discussion}.

In Figure \ref{figRealWorld} moving from top to bottom and left to right, we consider the following real world network examples:  the famous Zachary's Karate Club network which records social interactions of $n=34$ members of a karate club in the 1970s \citep{zachary_1977}; the canonical Florentine marriage network in which the $n=16$ nodes represent powerful families in Renaissance Florence and ties represent marital alliances \citep{padgett_1994, network};
%the well-known network of Sampson's monk data in which the $n=18$ nodes represent novices preparing to join a monastic order in New England and ties represent cumulative ``liking'' nominations \citep{sampson_1969, ergm};
Coleman's high school network in which the $n=73$ nodes represent boys in a small high school in Illinois during the fall of 1957 and ties are self-reported friendships \citep{coleman_1964, sna};
%an online social pet network where the $n=2426$ nodes represent members of the Hamsterster website and ties represent friendships or family links \citep{konect};
the neural network of the Caenorhabditis elegans (C.elegans) worm where the $n=297$ nodes represent neurons which are tied if at least one synapse or gap junction exists between them \citep{watts_strogatz_1998, tnet};
%a road network for the group of E-roads, located mostly in Europe, where the $n=1174$ nodes represent cities and ties represent a connection by an E-road \citep{subelj_bajec_2011,konect};
the road transportation network for Chicago where the $n=1467$ nodes represent traffic zones and ties represent links by freeways or other roads \citep{eash_chon_lee_etal_1979,konect};
and the power grid of the western U.S. where the $n=4941$ nodes represent either a generator, a transformator or a substation and ties represent a power supply line \citep{watts_strogatz_1998,konect}.  For each of these networks, plots of the eigenvalues in decreasing order suggest that the real world networks exhibit geometric features strongly reminiscent of hyperbolic space.  Further, for the non-social networks (see the networks in the right panel of Figure \ref{figRealWorld}), note that the eigenvalues for these networks are pushed even further towards the bottom left side of the plots.  This indicates that perhaps these networks require latent spaces that are more negatively curved or take better advantage of the negatively curved space (i.e., have a larger $R$\footnote{As mentioned in Section \ref{secStats}, many interesting features of hyperbolic space are more striking further away from the origin.  Recall that in the simulations for Figure \ref{figRealWorld}, we set $R=\pi$ across all spaces to make the latent spaces as comparable as possible; in elliptic space, $R$ is bounded above by $\pi$.  Increasing $R$, perhaps according to model \ref{hypModel.eq} or the approximate formula provided in \citet{krioukov_papadopoulos_kitsak_etal_2010}, could result in networks that better match the observed eigenvalues for these non-social networks.}).  See Appendix \ref{simModels.sec} for a discussion of how hyperbolic features manifest in networks simulated from common, intuitive models for network data.

%Furthermore, Weyl's Law and its generalizations strongly suggest (although do not directly imply) that the rate at which the eigenvalues increase should strongly correlate with the rate at which areas of disks grow as a function of radius.  Motivated by this line of thought, in the far right panel of Figure \ref{figLaPlacPic}, we also fit a quadratic curve to each of the network's series of eigenvalues (these curves are displayed in the center panel of Figure \ref{figLaPlacPic}) and compare the concavities, i.e. the second derivatives of the fitted curves. The distribution of these concavity parameters across the different latent space geometries more clearly delineates the role of geometry in network structure.
 
%%%%%%%%%%%%%%%%%%%%%%%%%%%%%%%%%%
%%
%%    (  6  )   DISCUSSION
%%
%%%%%%%%%%%%%%%%%%%%%%%%%%%%%%%%%%
\section{Discussion} \label{secDisc}

As we have demonstrated in this simulation exercise, using a negatively curved latent space allows us to smoothly grow the complexity of the latent space model \citep[i.e., accommodate more complicated network structure, as was expected from the intuitive arguments provided by][and as was demonstrated in Figure \ref{figDistBands}]{krioukov_papadopoulos_kitsak_etal_2010} without adding any additional structure to the model, such as explicit functions of network statistics as is done in ERGMs \citep{frank_strauss_1986, snijders_pattison_robins_2006}, or node-level random effects \citep{hoff_2005, krivitsky_handcock_raftery_hoff_2009}, or mixture distributions to spatially cluster the latent positions \citep{handcock_raftery_tantrum_2007}.  Furthermore, models in hyperbolic latent space (such as \ref{eqnKri} and \ref{simMod.eq}) preserve the use of distance to model dependence, a common feature of many popular statistical models, and is able to draw upon the long success of latent space models for network data. 
%Thus, we have provided evidence that using a hyperbolic latent space results in a novel statistical model for network data which preserves the simple form of the original latent distance model but is able to naturally model complex network structure, such as degree heterogeneity, as a simple consequence of the negative curvature of the hyperbolic latent space. 
We have provided an illustration of how a hyperbolic latent space can produce generated networks which align more closely to real-world networks in local and global structure (e.g. Figure \ref{figRealWorld}), as opposed to the popular choice of a Euclidean latent space.  A more thorough investigation to fully understand in which particular cases (i.e., under what kinds of assumptions about the network data generating mechanism) this holds is a promising avenue for future research.

The study of the types of graphs generated by latent space network models across different geometries, which we have provided here, should prove very helpful to practitioners in many fields.  Note that the viewpoint we have adopted here is commonly used in Bayesian analysis, for example in the assessment of posterior predictive distributions, where we are interested in understanding the behavior of a model through the data that it generates.  This view of network analysis can help improve practitioners' understanding of what kind of behavior to expect from different network models and, thus, which model may be most appropriate in a particular practical setting or application.  

Of course, many open questions about the relationship between geometric curvature and network properties remain unanswered.  For example, what is the precise role of the distribution for the latent positions in a geometric space of a particular curvature?  Future work might also consider continuing this investigation by considering a parametric family of spaces parametrized by their curvature, $\delta$:  from spaces of negative curvature ($\delta<0$) to Euclidean space ($\delta=0$) to spaces with positive curvature ($\delta>0$).  In this setting, we might imagine modeling observed networks by averaging over features of networks from both hyperbolic and elliptic space, for example.  We might also imagine performing inference for $\delta$ and using estimates of $\delta$ to describe properties of the observed network, or perhaps, to make comparisons across a set of observed networks.

Another exciting line of future research might consider a flipped perspective on the discussion provided here.  More specifically, given a particular set of network data can we use summary measures of observed network structure and/or graph Laplacians to infer the appropriate latent space geometry.  This type of approach might prove useful in model selection.

%%%%%%%%%%%%%%%%%%%%%%%%%%%%%%%%%%
%%
%%    APPENDIX
%%
%%%%%%%%%%%%%%%%%%%%%%%%%%%%%%%%%%

\appendix
\section{Models for visualizing hyperbolic space} \label{HypModels.sec}

\begin{figure}[h!]
  \centering
  \includegraphics[width=.5\textwidth]{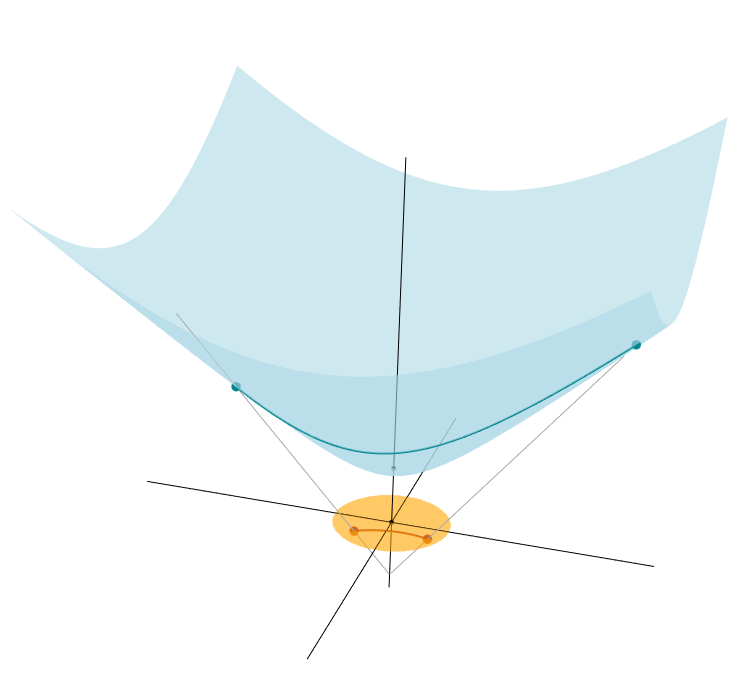}
  \caption{Visualizing Hyperbolic Space.  Turquoise points are plotted on the hyperboloid model (light blue), where the shortest distance between two points is a line across the surface of the hyperboloid.  Orange points are plotted on the Poincar\'{e} model, where the shortest distance betwen two points is the arc of the circle which intersects the two points and is perpendicular to the boundary.  Points in the Poincar\'{e} model are stereographic projections of points in the hyperboloid model - imagine sitting at $z=-1$ and looking through a disk of radius one, centered at $z=0$, up at the points on the hyperboloid (grey lines).}
  \label{figHypModels}
\end{figure}

 \begin{figure}[t]
%trim = left bottom right top
\includegraphics[width=1\textwidth]{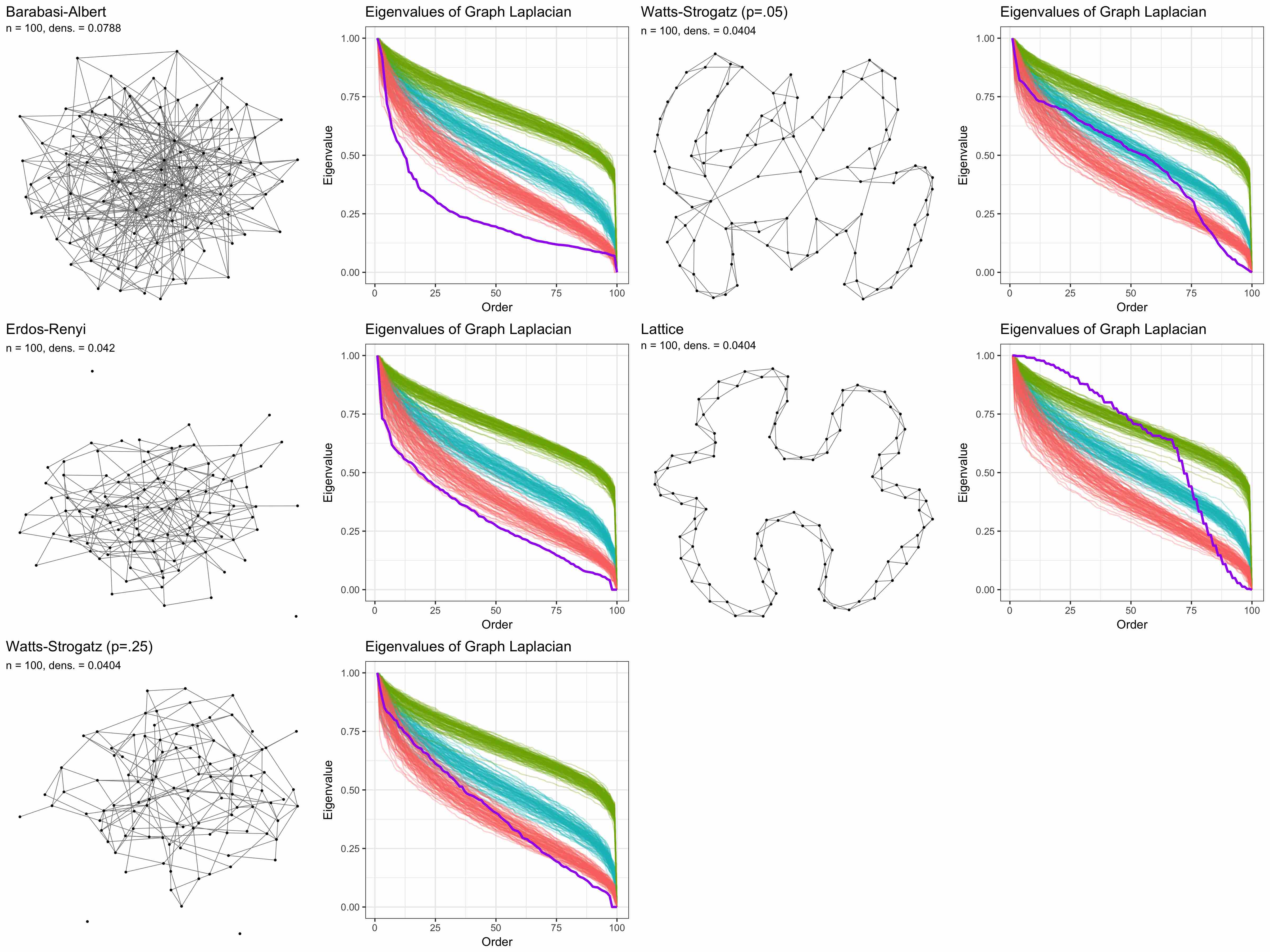}
\includegraphics[width=1\textwidth,trim={0 .5in 0 31in},clip]{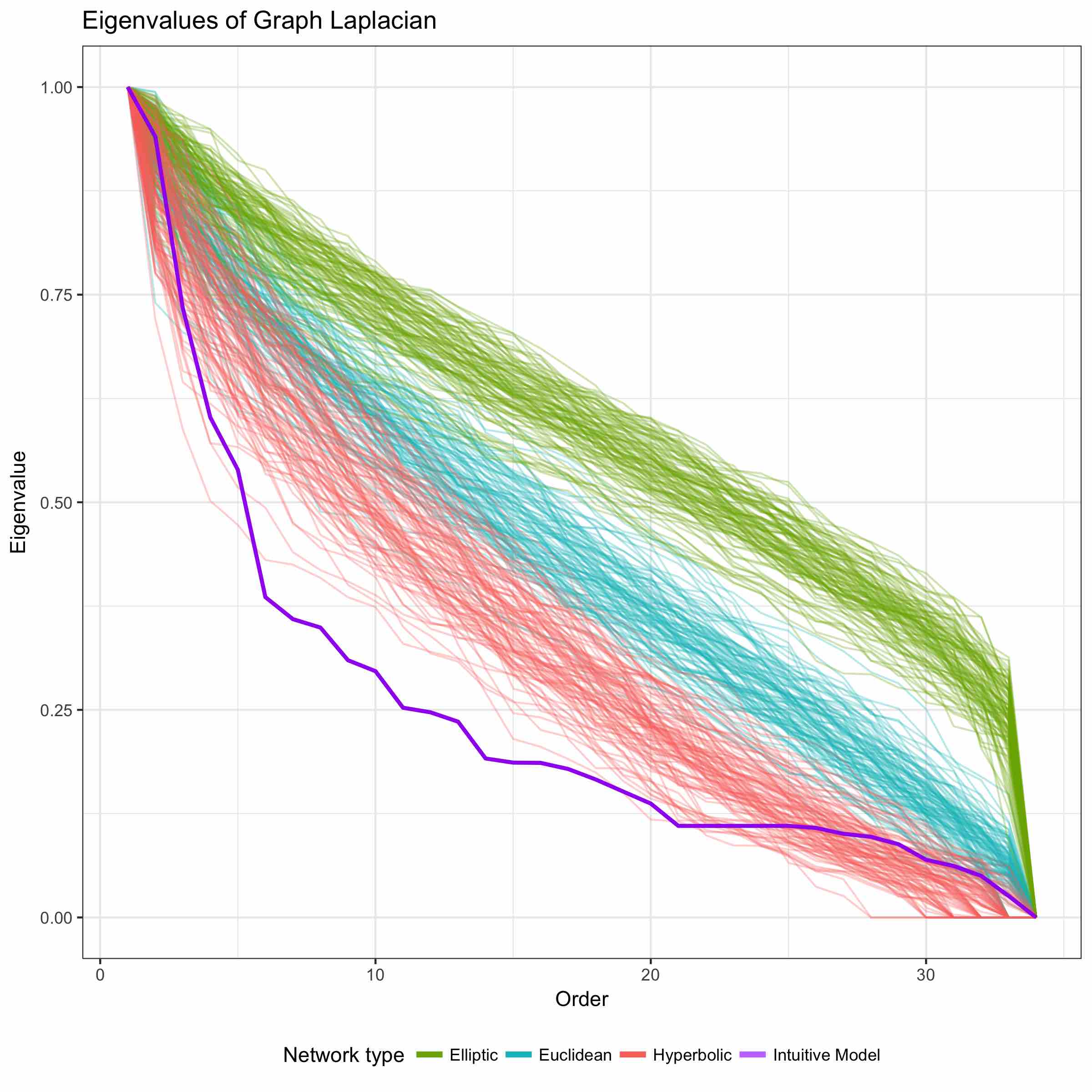}
\caption[Caption for LOF]{Intuitive models of network structure.  In these plots, networks simulated from intuitive models of network structure - the \BAmodel{} small world model, the \ERmodel{} model, the Watts-Strogatz model, and a perfect lattice - are compared to networks simulated from our general latent distance model (Equation \ref{eqLDMgen}) across different geometries. Each row includes a network plot and a plot of the eigenvalue curves, where the color corresponds to the geometry of the latent space.  In the eigenvalue plots, each network is represented by a curve which plots the network's eigenvalues against their order.   Parameters for the intuitive network models are specified such that each network has roughly the same density.}
\label{figSim}
\end{figure}

First, note that hyperbolic space is an example of a non-Euclidean space.  Recall that Euclidean space is described by Euclid's five postulates:
\begin{enumerate}
\item Each pair of points can be joined by one and only one straight line segment.
\item Any straight line segment can be indefinitely extended in either direction.
\item There is exactly one circle of any given radius with any given center.
\item All right angles are congruent to one another.
\item Given any straight line and a point not on it, there exists one and only one straight line which passes through that point and never intersects the first line, no matter how far they are extended.
\end{enumerate}
where the last postulate is referred to as the ``parallel postulate.''  Non-Euclidean spaces are formed by replacing the parallel postulate with some alternate behavior.  In hyperbolic space, it is replaced with the following:
\begin{enumerate}
\item[$5^*$.] Given any straight line and a point not on it, there are at least two distinct lines passing through that point which do not intersect with the line.
\end{enumerate}
Thus, in hyperbolic space, there exists infinitely many parallel lines (see Table \ref{tabCurve}).  In order to visualize the geometric properties of hyperbolic space, there are four popular analytic models:  the Beltrami-Klein model, the Poincar\'{e} disk model, the Poincar\'{e} half-plane model, and the hyperboloid model (although other models certainly exist).  Each of these models define a hyperbolic plane which satisfies the axioms of a hyperbolic geometry (1,2,3,4,and $5^*$).  Moving forward, we will primarily rely on the hyperboloid model to visualize positions in hyperbolic space but will sometimes refer to the Poincar\'{e} disk model as well.

\textbf{Hyperboloid model}.  Although specified for $n$-dimensional hyperbolic space, we will typically focus on a $2$-dimensional hyperbolic space, $\mathbb{H}^2$, and describe this version of the model here.  Points on the hyperbolic plane are modeled as points on the surface of the upper (or lower) sheet of a hyperboloid in $3$-dimensional Euclidean space, $\mathbb{R}^3$ (consider the blue points in Figure \ref{figHypModels}).  That is, all points $(x,y,z)$ in the plane satisfy the following
\[ z^2 - y^2 - x^2 = 1, \]
with $z>0$.  Distance between points in the hyperboloid model is simply distance along the surface of the hyperboloid; this mimics visualization of distance in elliptic space, as the distance (arc length) along the surface of the hypersphere.

\textbf{Poincar\'{e} disk model}.  Unlike the hyperboloid model, the Poincar\'{e} disk model represents (infinite) hyperbolic space within a finite space.  This model is formed by the stereographic projection of the hyperboloid in $\mathbb{R}^3$ onto the unit disc at $z=0$.  As points in this model move closer to the boundary of the disk, they approach inifinity; in the hyperboloid model, this is equivalent to moving further up the surface of the hyperboloid (consider the orange points in Figure \ref{figHypModels}).  Distances in this model are semicircles perpendicular to the boundary of the Poincar\'{e} disk.

%%%%%%%%%%%%%%%%%%%%%%%%%%%%%%%%%%
%%
%%    APPENDIX
%%
%%%%%%%%%%%%%%%%%%%%%%%%%%%%%%%%%%

\section{Hyperbolic features in intuitive network models} \label{simModels.sec}

Perhaps non-CLS models for network data also incorporate some of the same features we observed in the networks simulated in a hyperbolic space.  In order to examine this idea, we replicate the analysis used to create Figure \ref{figRealWorld} with networks simulated from common intuitive models for networks.  Here, we examine the \BAmodel{} model \citep{barabasi_albert_1999}, the \ERmodel{} model \citep{erdos_renyi_1960}, the Watts-Strogatz model \citep{watts_strogatz_1998} and a perfect lattice\footnote{For the \BAmodel{} model, the number of edges added in each time step is 4 (one time step is used to simulate the network); for the \ERmodel{} model, the probability of a tie is 0.08080808; for the Watts-Strogatz model, a neighborhood of 2 vertices of the lattice are initially connected and the rewiring probability is set to 0.25 and 0.05, respectively.  All networks are simulated with $n=100$ nodes using standard functions in the \texttt{igraph} package for \texttt{R} and plotted using the default settings of the \texttt{ggnet} function in the \texttt{GGally} package for \texttt{R}.}.  The eigenvalue plots indicate that many of these network models capture the same kind of features we observe for networks in a hyperbolic space, particularly for models that create highly heterogenous networks (like the \BAmodel{} model) or highly random networks (like the \ERmodel{} model and the Watts-Strogatz model with larger rewiring probability).  For example, the \BAmodel{} model uses a preferential attachment mechanism to generate scale-free networks that are characterized by highly right-skewed degree distributions, as we observed in hyperbolic space in Figure \ref{figDistHist}.  However, the perfect lattice and some variations of the Watts-Strogatz model (which modifies a perfect lattice with random rewirings) provide nice nonexamples;  these highly regular networks, which are neither very heterogeneous nor very random \citep[see Figure 3 in][for more discussion]{sole_valverde_2004}, do not exhibit hyperbolic features.

%%%%%%%%%%%%%%%%%%%%%%%%%%%%%%%%%%
%%
%%    Acknowledgements
%%
%%%%%%%%%%%%%%%%%%%%%%%%%%%%%%%%%%
\section*{Acknowledgements}
Smith and Calder were partially supported by grants from the National Science Foundation (NSF DMS-1209161), the National Institutes of Health (NIH R01 HD088545), and The Ohio State University Institute for Population Research (NIH P2CHD058484).

%%%%%%%%%%%%%%%%%%%%%%%%%%%%%%%%%%
%%
%%    BIBLIOGRAPHY
%%
%%%%%%%%%%%%%%%%%%%%%%%%%%%%%%%%%%
\bibliography{../../../BibStuff/compare}
\end{document}